\documentclass{iopjournal}

\usepackage[acronym]{glossaries}
\glsdisablehyper
\makeglossaries
\newacronym{mastu}{MAST-U}{Mega Ampere Spherical Tokamak Upgrade}
\newacronym{tbr}{TBR}{tritium breeding ratio}
\newacronym{dt}{DT}{deuterium-tritium}
\newacronym{dd}{DD}{deuterium-deuterium}
\newacronym{csg}{CSG}{constructive solid geometry}
\newacronym{cad}{CAD}{computer-aided design}
\newacronym{dagmc}{DAGMC}{Direct Accelerated Geometry Monte Carlo}
\newacronym{fendl}{FENDL}{Fusion Evaluated Nuclear Data Library}
\newacronym{endfb}{ENDFB}{Evaluated Nuclear Data File/B-VII}
\newacronym{jeff}{JEFF}{Joint Evaluated Fission and Fusion}
\newacronym{tf}{TF}{toroidal field}
\newacronym{pf}{PF}{poloidal field}
\newacronym{cs}{CS}{central solenoid}
\newacronym{lcfs}{LCFS}{last closed flux surface}
\newacronym{mhd}{MHD}{magnetohydrodynamics}
\newacronym{st}{ST}{spherical tokamak}
\newacronym{fw}{FW}{first wall}
\newacronym{sdr}{SDR}{Shutdown dose rate}
\newacronym{pfc}{PFC}{plasma facing components}
\newacronym{mpi}{MPI}{message passing interface}

\newacronym{gs}{GS}{Grad-Shafranov}
\newacronym{vsc}{VSC}{vertical stability coil}

\usepackage[version=4]{mhchem} % Chemical formulae / isotopes

\usepackage{siunitx} % Units

\usepackage{enumitem}

\usepackage{caption}

\usepackage{subcaption}

\usepackage{tabularx}

\usepackage[backend=bibtex, sorting=none, mcite=true]{biblatex}
\newcommand{\jg}{\textsc{Jenga}}

\begin{document}

\articletype{Paper} %	 e.g. Paper, Letter, Topical Review...

% \title{MAST-U study on JENGA}
% \title{Jenga framework for integrated design and modeling of Tokamaks: A case study on MAST-U}
\title{Magnetohydrodynamic equilibrium and neutronics study on MAST-U using \jg\ framework}

% \author{Author Name$^1$\orcid{0000-0000-0000-0000}, Author Name$^2$\orcid{0000-0000-0000-0000} and Author Name$^{1,*}$\orcid{0000-0000-0000-0000}}
\author{Saptarshi Rajan Sarkar$^1$\orcid{0009-0002-7703-3647}, Rahul Babu Koneru$^1$\orcid{0000-0002-5212-1360}, 
Ravi Gupta$^1$, 
Roshan George$^1$, Animesh Kuley${^2}$, Santosh Ansumali${^3}$\orcid{0000-0002-6938-233X} and Shaurya Kaushal$^{1,*}$\orcid{0000-0002-0488-2413}}

\affil{$^1$Pranos Fusion Private Limited,  Innovation and development centre,
Bengaluru, India}

\affil{$^2$Department of Physics, Indian Institute of Science, Bangalore 560012, India}

\affil{$^3$Engineering Mechanics Unit, Jawaharlal Nehru Centre for Advanced Scientific Research, Bengaluru, India}

\affil{$^*$Author to whom any correspondence should be addressed.}

\email{shaurya@pranosfusion.energy}


\begin{abstract}
Tokamak design is inherently challenging due to several cross-competing effects which require a careful and calibrated treatment to obtain an optimal operational envelope. Incorporating physics across varied fidelities is crucial in this exercise. \jg\ is developed as a unified design and modeling framework for tokamaks, seamlessly coupling systems-level studies to high-fidelity models based on first principles. In this work, static \gls{gs} equilibrium for an entire pulse and the neutronics study of the \gls{mastu} tokamak are carried out in \jg. Coil currents and plasma profiles from the EFIT++ reconstruction of \gls{mastu} shots are used to reproduce the plasma poloidal flux and shape targets at different time slices. The results from \jg\ are also in good agreement with FreeGSNKE and Fiesta codes. Neutronics analysis is performed for a hypothetical 50-50 mixture of \gls{dt} fuel, using the same data structure as the systems and equilibrium studies. A distributed neutron source is initialized within the \gls{lcfs} of the plasma, with their strength being functions of the density and temperature of the ions. The distribution of the neutron flux across the energy spectrum is computed for the active coils and the first wall (limiter) independently over multiple scenarios. We demonstrate the capabilities of \jg\ with a comprehensive analysis that takes inputs about the plasma geometry, tokamak design and plasma profiles and performs 0D, 2D and 3D numerics for the systems study, equilibrium and neutron transport respectively.
\end{abstract}

\section{Introduction} 
\label{sec:intro}
The design and operation of a tokamak is fundamentally an optimization problem
with constraints set on several engineering and operational parameters such as
plasma shape, plasma current, toroidal magnetic field strength, heating power and confinement time among others. Each one of these parameters constrains the other through coupled physics and engineering limits necessitating an integrated treatment to
identify which parameter combinations are simultaneously achievable. This
coupling motivates systems-level codes which self-consistently connect plasma
physics, electromagnetics and engineering constraints within a unified
computational framework. Several such codes have been developed and have proven
their value in the context of fusion device design. PROCESS~\cite{kovari20143054,kovari20169} and SYCOMORE~\cite{reux2015demo} and STOP~\cite{mccool1994} are few examples that serve this role for
large-aspect-ratio and spherical tokamaks. GLOBSYS extended this
zero-dimension (0D) approach to Globus-M and performed systematic benchmarking
across NSTX, NSTX-U, MAST, MAST-U, and ST40~\cite{mineev2022engineering}.

% These systems codes are primarily based on scaling laws and empirical relations
A complete picture of tokamak operations requires moving between 0D systems
codes, one-dimensional (1D) transport, two-dimensional (2D) equilibrium solvers
and three dimensional (3D) magnetohydrodynamics or neutronics codes, each with
their own input format, conventions, and machine-specific setup. In practice, this
means that results at one level of fidelity are not automatically consistent
with results at another, and every new machine requires bespoke setup work at
each level independently. Recently developed integrated frameworks such as JINTRAC~\cite{romanelli2014}, BLUEMIRA~\cite{coleman2019blueprint} and FUSE
\cite{meneghini2024fuse} offer some of the aforementioned capabilities for
self-consistent reactor modeling with consistent data structures.

\begin{figure}[hbt!]
     \centering
     \includegraphics[width=0.8\textwidth]{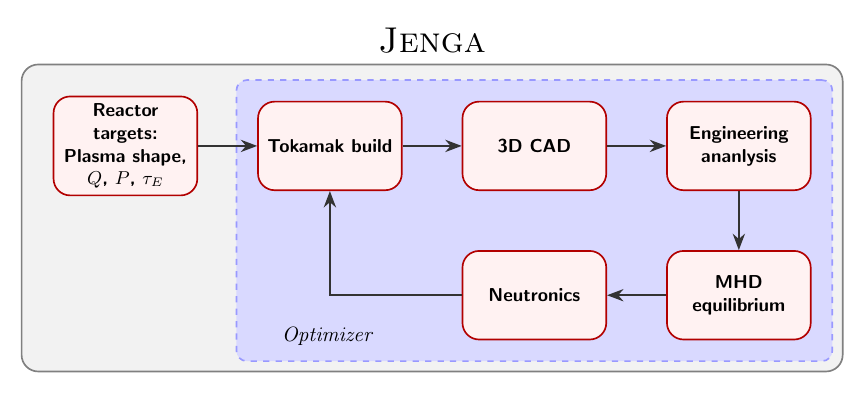}
     \caption{The operational flow diagram of \jg.}
     \label{fig:jenga_flow}
\end{figure}

\jg\ provides a single integrated platform spanning 0D systems-level analysis,
2D plasma equilibrium, 3D engineering analysis (structural and electromagnetic
loads) and neutronics all encompassed under an optimization layer. \jg\
seamlessly connects all these different blocks to provide an integrated
environment with an improved fidelity for informed decision making. 
This is made possible by a uniform data structure with machine specific inputs
like geometry, coil configuration, material properties. This single description
then propagates consistently through every level of the analysis. The framework
is thereby machine-agnostic by construction. A high-level
overview of \jg\ is shown in Figure \ref{fig:jenga_flow}. For a given set of fusion
reactor requirements, a radial build can be obtained
following a self-consistent systems-level analysis. Optionally, the radial build can be further optimized for any given set of design
parameters and constraints. The radial build can then be passed on to build a 3D
CAD model with options to add ports, support structures for coils, cryostat and vacuum vessel.
A comprehensive structural and electromagnetic analysis can be performed on this
3D model to assess the effects of various loads such as the self-weight, vacuum pressure
and electromagnetic forces. The position and size of the \gls{pf}
coils can be optimized for the target plasma parameters using an inverse
\gls{gs} solve. On the other hand, a forward solve can be used to
prototype actuator control systems and pulse waveforms. This acts as a `flight
simulator' for the virtual tokamak. Furthermore, detailed neutronics study can
be carried out to evaluate neutron flux on the different tokamak structures such
as the first wall, the blanket and the vacuum vessel. If necessary, results from the neutronics analysis can be used to update the radial build forming a closed-loop iterative development cycle. 

% Onboarding a new machine consists of populating this machine description schema
% with machine specific inputs like geometry, coil configuration, material
% properties. That single description then propagates consistently through every
% level of the analysis. The framework is thereby machine-agnostic by
% construction.

This paper presents the first application of \jg\ using MAST-U as the validation
target. MAST-U is a well-suited choice for several reasons. Its shot database is
publicly archived and includes equilibrium reconstructions from EFIT++
\cite{Lao_1985,Lao01102005}, providing ground-truth flux surface geometry for
direct comparison. Prior independent equilibrium studies
\cite{pentland2025validation} using Fiesta \cite{cunningham20133238} and FreeGSNKE \cite{amorisco2024} provide comparison
baselines for a code-to-code assessment of the equilibrium solver in \jg. The
machine also presents non-trivial geometric complexity in the form of its
Super-X divertor configuration, which represents a demanding test of flux
surface resolution beyond the conventional single-null case. The equilibrium
solver integrated into \jg\ for this study is ToKaMaker \cite{hansen2024109111},
an open-source finite element solver for the time-dependent Grad-Shafranov
equation. We demonstrate forward equilibrium solutions for both a conventional
divertor configuration (shot 45425) and a Super-X divertor configuration (shot
45292). A detailed analysis is carried out comparing the computed separatrix geometry, shape targets and flux measurements against the EFIT++
reconstructions in conjuction with the Fiesta and FreeGSNKE codes for the entire shot. We additionally present the neutronics module integrated through
OpenMC \cite{romano2015}, establishing the workflow to estimate neutron flux
from the computed equilibrium geometry. The neutron flux distribution is computed in the first wall and the active coils. We also show feedback capabilities of our framework, where we can make impromptu build modifications through the \gls{cad} and compute the resultant neutron flux per layer, which provides essential design metrics.

% The plasma
% control system (PCS) stack of Jenga is out of scope for this paper
% and will be presented in a subsequent publication. The present work focuses on
% the simulation and benchmarking layer: demonstrating that MAST-U onboarding
% requires only machine-specific inputs to a general framework, with no
% machine-specific code written, and that the resulting forward solutions are in
% quantitative agreement with the experimental reconstructions.

The paper is organized as follows. The description of the MAST-U tokamak is presented in Section \ref{ssec:mastu_description} followed by an introduction to the \jg\ software and running MAST-U on \jg\ in Section \ref{ssec:jenga}. Next, the \gls{gs} equation is briefly discussed in \ref{sec:gs_eq}. The forward solve results of the MAST-U shots from TokaMaker and a comparative analysis of these results with Fiesta and FreeGSNKE is presented in \ref{sec:forward_solve}. A detailed description the neutronics workflow using OpenMC and the neutron flux calculations on the MAST-U tokamak are presented in Section \ref{sec:neutronics}.

% MAST-U description
\section{\gls{mastu} on \jg} \label{sec:mastu_jenga}

\subsection{Description of MAST-U geometry} \label{ssec:mastu_description}
\begin{figure}
 \centering
        \includegraphics[width=0.5\textwidth]{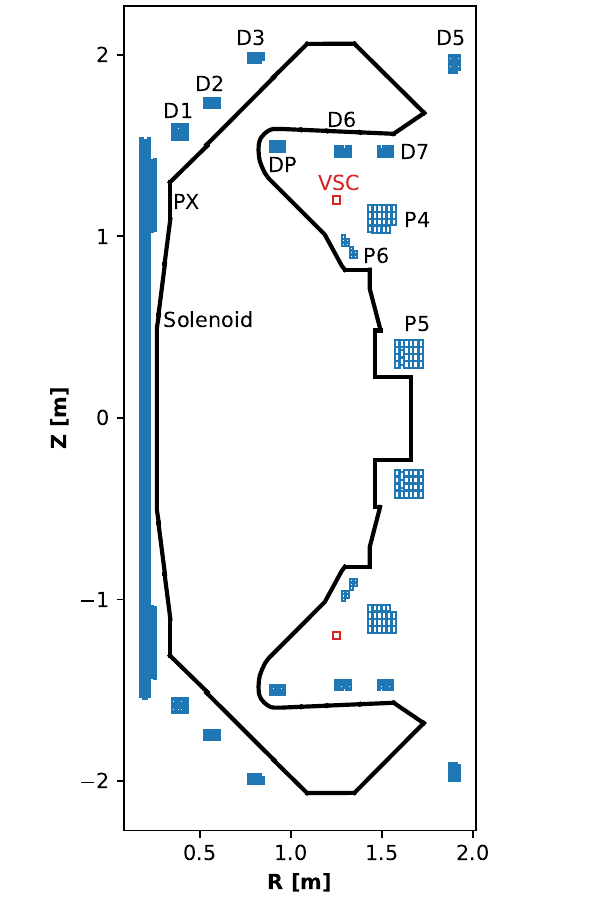}
 \caption{The poloidal cross-section of the MAST-U tokamak with all the active coils (blue) and limiter (black) is shown. The artificial vertical stability coil is shown in red.}
\label{fig:machine}
\end{figure}

% \subsection{Active coils}
The poloidal cross-section of the MAST-U tokamak used in the TokaMaker
simulations is shown in Figure \ref{fig:machine}. It consists of 12 pairs of
active \gls{pf} coils (shown in blue) which are used for plasma generation, shaping and
control. Each of these coils is in turn composed of multiple rectangular
filaments. The function of each of these coils is mentioned in Table
\ref{tab:mastu_coils}. An additional pair \gls{vsc} (shown in red) is added to
provide vertical stability to the plasma. This is an artificial coil pair whose
necessity is discussed in detail in Section \ref{sec:forward_solve}.

% \subsection{Limiter}
The limiter is composed 98 pairs of $(R,Z)$ coordinates describing the polygon.
The limiter is shown in black in Figure \ref{fig:machine}. The plasma is
confined entirely within the limiter. The passive structures such as the coil
cases, vacuum vessel, center column, support structures etc. are excluded in
this work since the induced currents in these structures are not used in the
static \gls{gs} solve.

\begin{table}
\caption{The different active coils in the MAST-U tokamak and their function.}
\centering
\begin{tabular}{l l}
\hline
Coil name & Function \\
\hline
P1 & Solenoid for inductive startup \\
P4/P5 & Radial position control and shape control \\
P6 & Vertical position control \\
D1/D2/D3/PX & X-point position and divertor leg control \\
DP & Additional X-point control and flux expansion \\
D5 & Super-X leg radius control \\
D6/D7 & Expansion control \\
\hline
\end{tabular}
\label{tab:mastu_coils}
\end{table}

\subsection{Configuration} \label{ssec:jenga}
Configuring MAST-U on \jg\ requires two classes of inputs:
\begin{enumerate}[label=(\roman*)]
    \item the machine description, which is fixed for the lifetime of the machine and
    \item the shot-specific inputs, which specify the plasma and coil state at a given time.
\end{enumerate}

\begin{figure}[hbt!]
     \centering
     \includegraphics[width=0.9\textwidth]{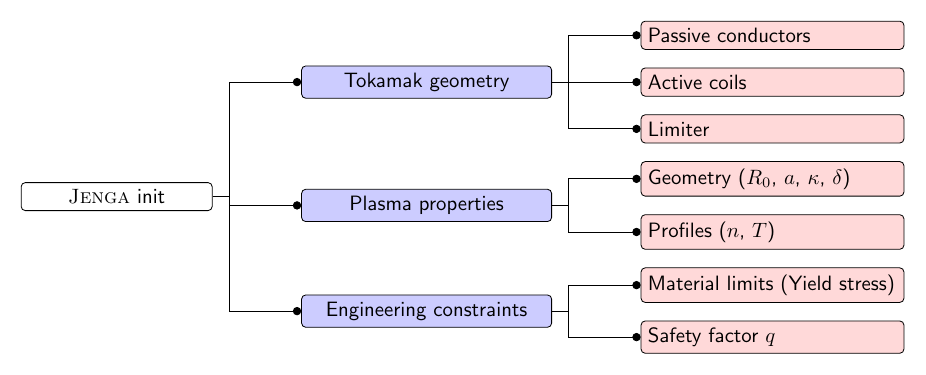}
     \caption{A non-exhaustive list of inputs required during \jg\ initialization. The plasma geometry is comprised of the plasma major radius $R_0$, minor radius $a$, elongation $\kappa$ and triangularity $\delta$. The plasma profiles correspond to the number density $n$ and temperature $T$ of the species.}
     \label{fig:jenga_setup}
\end{figure}

A non-exhaustive list of \jg\ inputs are shown in Figure \ref{fig:jenga_setup}.
For an existing tokamak, the tokamak geometry can be readily provided. In the
case of a conceptual tokamak, this will depend on a closed-loop optimization of
active coil positions and sizes subject to a target plasma shape. A target
plasma geometry acts as a starting point for a systems-level study to identify
the probable zones of tokamak operations subject to engineering constraints such
as the material stress limits, proximity limits of coils, force limits on the
supporting structures and  safety factor. Additionally, currents in active and
passive conductors, \gls{lcfs} location, X-points, strikepoints and poloidal
flux on \gls{lcfs} can also be provided depending on the mode of \jg\ operation.

For setting up MAST-U, the EFIT++ reconstruction data from UKAEA
\cite{pentland2025validation} saved as python pickle files is used. The tokamak
geometry, coil currents, shape targets etc. for the entire pulse are stored as
python dictionaries in the pickle files. This information is then used to
generate the geometries required for the \gls{mhd} and the neutronics study. For
the \gls{mhd} study, the coil currents, plasma profiles and the magnetic axis
are additionally used. Whereas in the case of neutronics, the 3D geometry is
constructed using the Cadquery backend.

\section{Magnetohydrodynamic governing equations} \label{sec:gs_eq}
% \subsection{Governing equations}
% \subsection{Grad-Shafranov Equation}
The Grad-Shafranov equation describes a plasma under static \gls{mhd} equilibrium wherein the plasma pressure is balanced by external magnetic field. This is a non-linear elliptic partial differential equation derived in axisymmetric cylindrical coordinates and is given by
\begin{equation} \label{eq:gs-full}
       R \frac{\partial}{\partial R} \left(\frac{1}{R}
       \frac{\partial \psi}{\partial R} \right) 
       + \frac{\partial^2 \psi}{\partial Z^2} 
       %= -\mu_0 R^2 p^{\prime} - ff^{\prime}
       = -\mu_0 R J_{\phi}(R,Z) 
\end{equation} 
where $ \psi $ is the poloidal flux function (V.s/rad), $J_{\phi}$ is the toroidal current density ($\mathrm{A/m^2}$) and $\mu_0$ (H/m) is the permeability of free space. The toroidal current density $J_{\phi}$ is the sum of current density contributions from the plasma $J_{P}$ and the current carrying conductors $J_{C}$. 

Within the plasma region, $J_{P}$ is non-zero and is given by
\begin{equation}
    J_{P} = R \frac{dp}{d \psi}  + \frac{1}{\mu_0 R}F(\psi)\frac{dF}{d \psi}
\end{equation}
where $p$ is the isotropic plasma pressure (Pa) and $F$ is the toroidal field function defined as $F=R B_{\phi}$ (T.m) where $B_{\phi}$ is the toroidal magnetic field. The current density in the $i-th$ discrete conductor is given by 
\begin{equation}
    J_{C_{i}} = \frac{I_{C_{i}}}{\int_{C_{i}} dA}
\end{equation}
where, $I_{C_{i}}$ is the current in the conductor and the integral in the denominator is area of the conductor.

There are typically two types of boundary conditions for the Grad-Shafranov equation. The first is a fixed-boundary condition wherein the flux on the boundary is set to a constant. The second is a free-boundary condition where the flux at the boundary is ensured to be consistent with flux generated from the plasma as well as the coils and is not available a priori. An efficient way to compute the flux on the boundary is the van Hagenow's method \cite{jardin2010} which relates the magnetic field in a region to the surface integral of the tangential field on that region. The flux on the boundary is then given by
\begin{equation} \label{eq:gs_bc}
    \psi(\boldsymbol{r_b}) = \oint \frac{1}{R} G(\boldsymbol{r_b};\boldsymbol{r}) 
                             \frac{\partial U}{\partial n} dl 
                              + \int_C G(\boldsymbol{r_b};\boldsymbol{r}) J_C dA
\end{equation}
where $\boldsymbol{r_b}$ is the position ($R_b, Z_b$) of the boundary nodes. The first term on the right hand side is the contribution from the plasma current while the second term is the contribution from the external (conductor) currents. The toroidal Green's function $G(\boldsymbol{r_b};\boldsymbol{r})$ is given by 
\begin{equation} \label{eq:green_func}
    G(\boldsymbol{r_b};\boldsymbol{r}) = \frac{\mu_0}{2\pi} \frac{\sqrt{R R_b}}{k} [(2-k^2)K(k)-2E(k)],
\end{equation} 
where, $E(k)$ and $K(k)$ are the complete elliptic integrals of the first and second kind respectively with 
\begin{equation} \label{eq:green_func_k}
    k^2 = \frac{4R R_b}{(R+R_b)^2+(Z-Z_b)^2}.
\end{equation}
The function $U$ in Equation \ref{eq:gs_bc} is the solution of Equation \ref{eq:gs-full} with homogenous boundary condition $U=0$ on the boundary.
% \subsection{Ideal MHD stability equations}

\section{Forward Solve} \label{sec:forward_solve}
The finite element solver TokaMaker \cite{hansen2024109111} is used to compute the plasma equilibrium using a forward solve wherein the plasma equilibrium is computed for a given set of coil currents and plasma current density profile. Due to the non-linear nature of the \gls{gs} equation, a fixed-point iteration (Picard) is used to solve for $\psi$. It is well-known that fixed-point iterations in a forward \gls{gs} solve are inherently unstable due to a combination of physical and numerical instabilities \cite{carpanese2021development}. To counteract this instability, a fictitious pair of \gls{vsc} is added close to the P6 coil, which is the actual \gls{vsc}. 

Data from the EFIT++ reconstruction of the MAST-U discharges are used to compute the plasma equilibrium in a conventional (shot 45425) and a Super-X (shot 45292) divertor configuration. This data is comprised of plasma current $I_p$, magnetic axis $(R_0,Z_0)$, pressure at the axis $p_0$, profiles $p'$, $FF`$, currents in active and passive structures, diagnostic measurements among other quantities. In the present work, $I_p$ and the magnetic axis, $p'$, $FF`$, and the active coil currents are inputs to TokaMaker. With the addition of the \gls{vsc}, the coil currents in TokaMaker can be matched exactly to that of EFIT++ and a fair comparison can be made with the published results \cite{pentland2025validation}.

\begin{figure}[hbt!]
     \centering
     \includegraphics[width=1.0\textwidth]{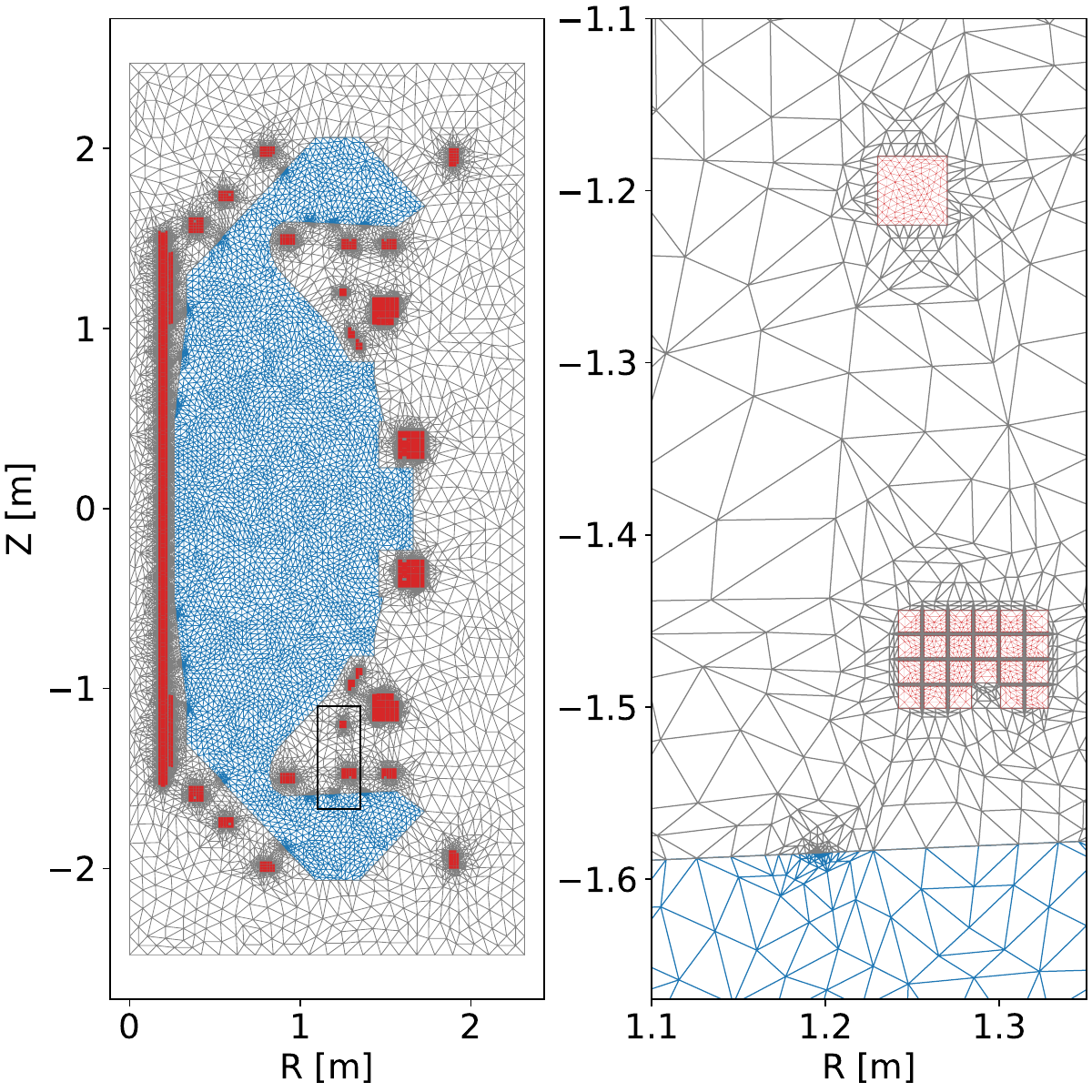}
     \caption{Mesh used in TokaMaker simulations (left panel) depicting the plasma region (blue), vacuum region (gray) and the coil region (red). A close-up of the region denoted by the black rectangle is shown in the right panel.}
     \label{fig:dev_mesh}
 \end{figure}

\begin{figure}[hbt!]
     \centering
     \includegraphics[width=0.7\textwidth]{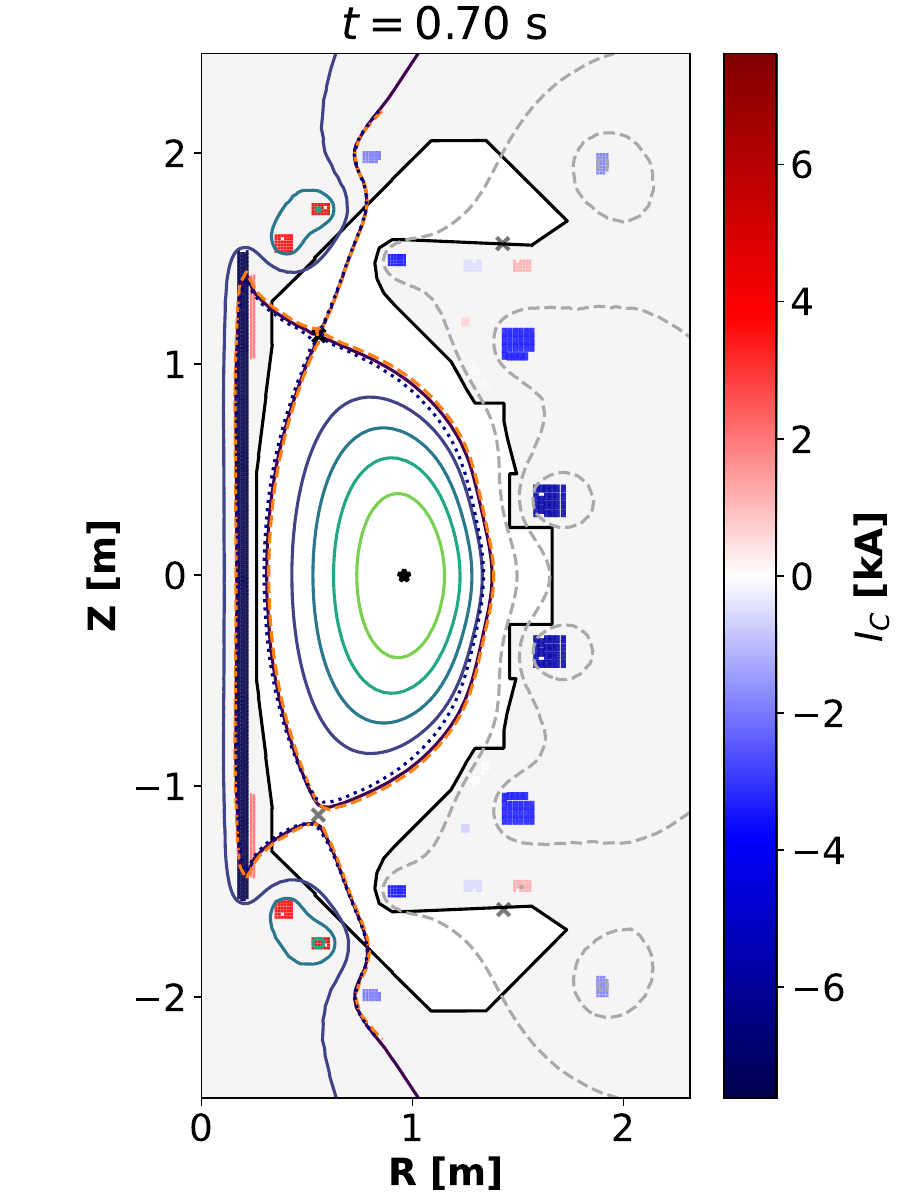}
     \caption{Poloidal flux contours at equilibrium computed by TokaMaker for shot 45425. The separatrix from Fiesta (dashed orange) and FreeGSNKE (dotted blue) are overlaid on to the TokaMaker solution.}
     \label{fig:psi_comparision}
 \end{figure}

The computational domain in TokaMaker extends from 0.0 m to 2.4 m in the radial ($R$) direction and from -2.7 m to 2.7 m in the vertical ($Z$) direction. The computational grid is made up of triangular elements and is classified in to vacuum, plasma and coil regions. The grid in the different regions is shown in Figure \ref{fig:dev_mesh}. The EFIT++ reconstruction was carried out on a $65 \times 65$ rectangular grid with $R \in [0.06,2.0]$ and $Z \in [-2.2,2.2]$ which is the same for Fiesta and FreeGSNKE. While TokaMaker allows for up to a fourth-order polynomial basis functions, we use a second-order polynomial basis for the all the simulations reported in this paper. 
A snapshot of plasma equilibrium in a conventional divertor at $t=0.7\ s$ and the corresponding mesh are shown in Figure \ref{fig:psi_comparision}. The normalized poloidal flux ($\psi$) contours from TokaMaker's solution is shown along with the separatrices from Fiesta, FreeGSNKE and EFIT++ overlaid on top. As seen from the figure the differences are quite small. The absolute difference in flux on the axis $\psi_a$ and the boundary $\psi_b$ is between $10^{-3}$ Wb and $10^{-2}$ Wb between EFIT++ and the three \gls{gs} solvers. 
% One distinction here is that, the passive structures are not modelled in TokaMaker
 
% \begin{figure}
%  \centering
%  \begin{subfigure}{0.6\textwidth}
%      \centering
%      \includegraphics[width=1.0\textwidth]{figures/dev_mesh_v2.pdf}
%      \caption{Poloidal flux contours at equilibrium computed by TokaMaker for shot 45425. The separatrix from Fiesta (dashed orange) and FreeGSNKE (dotted blue) are overlaid on to the TokaMaker solution.}
%      \label{fig:psi_comparision}
%  \end{subfigure}
%  \begin{subfigure}{0.4\textwidth}
%      \centering
%      \includegraphics[width=1.0\textwidth]{figures/psi_comparision_oft_fiesta_nke.pdf}
%      \caption{Poloidal flux contours at equilibrium computed by TokaMaker for shot 45425. The separatrix from Fiesta (dashed orange) and FreeGSNKE (dotted blue) are overlaid on to the TokaMaker solution.}
%      \label{fig:psi_comparision}
%  \end{subfigure}
% \end{figure}

\subsection{Conventional Divertor}
In the conventional divertor configuration, the plasma at the flat-top has a double-null shape with an approximate plasma current of 750 kA. The plasma is heated using two neutral beam injection (NBI) systems and remains in H-mode confinement for a majority of the shot.

\begin{figure}[hbt!]
 \centering
        \includegraphics[width=1.0\textwidth]{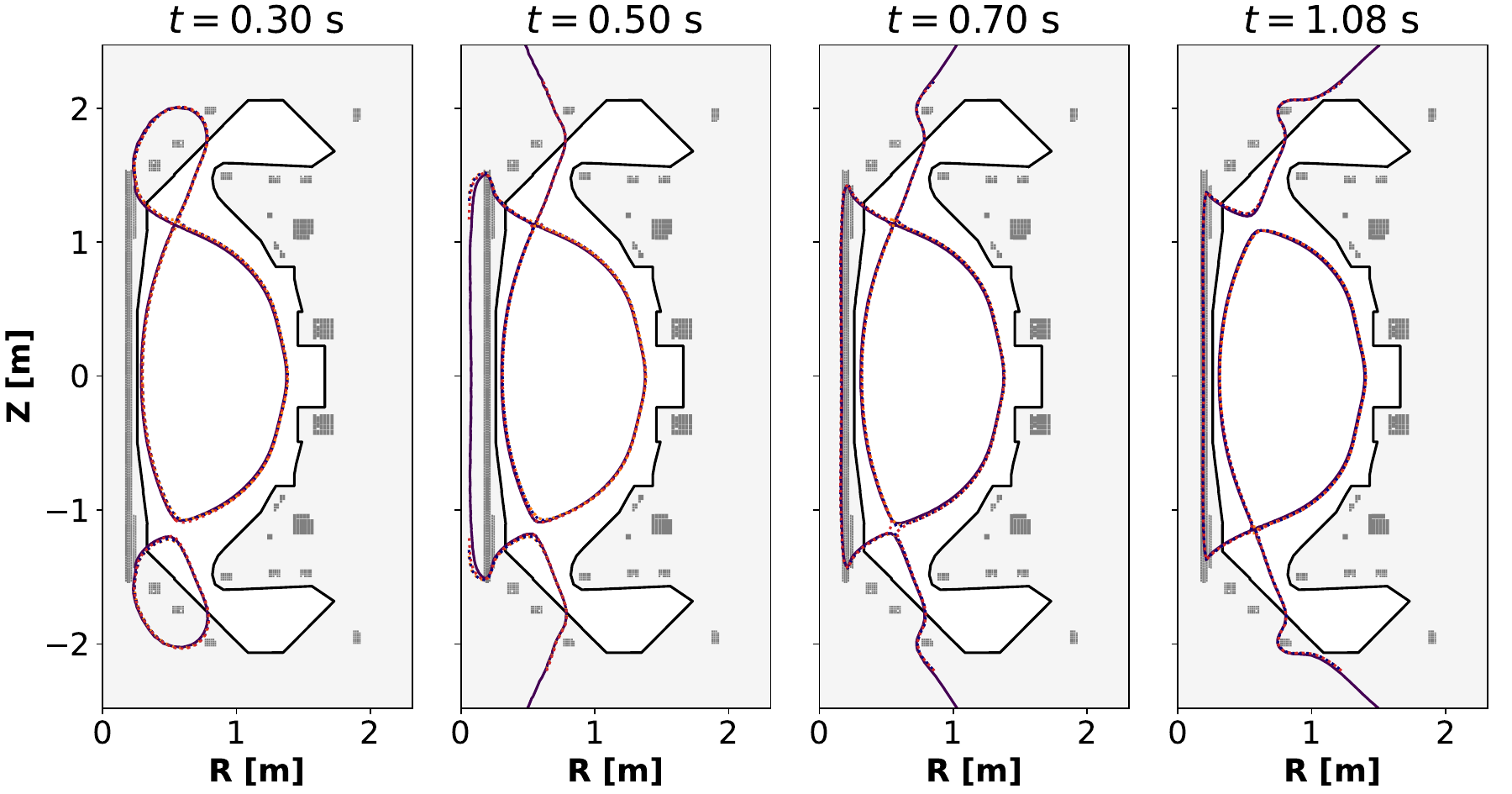}
 \caption{The evolution of TokaMaker (solid black), Fiesta (dotted orange), FreeGSNKE (dotted blue) and EFIT++ (dotted red) separatrices at different times in the conventional divertor configuration.}
\label{fig:lcfs_conv}
\end{figure}

\begin{figure}[hbt!]
 \centering
        \includegraphics[width=0.8\textwidth]{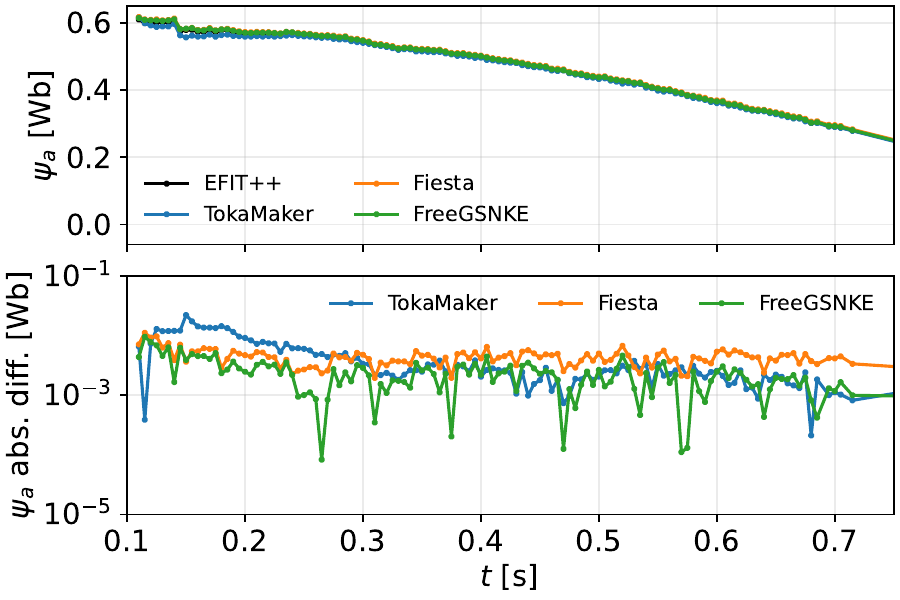}
 \caption{The evolution of poloidal flux at the magnetic axis of the plasma (top panel) and the absolute difference between EFIT++ and TokaMaker, Fiesta and FreeGSNKE (bottom panel) in the conventional divertor configuration.}
\label{fig:psia_conv}
\end{figure}

\begin{figure}[hbt!]
 \centering
        \includegraphics[width=0.8\textwidth]{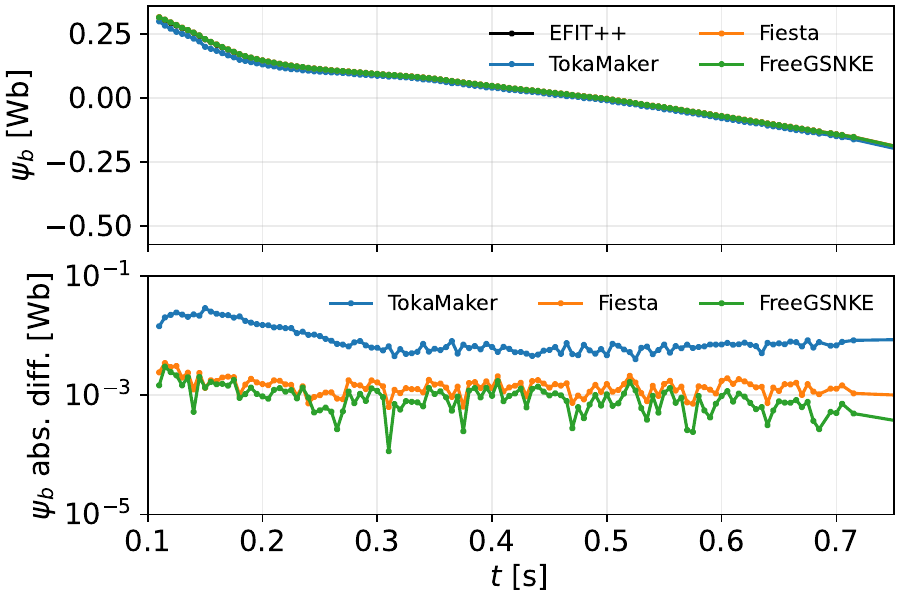}
 \caption{The evolution of poloidal flux at the plasma boundary (top panel) and the absolute difference between EFIT++ and TokaMaker, Fiesta and FreeGSNKE (bottom panel) in the conventional divertor configuration.}
\label{fig:psib_conv}
\end{figure}

The evolution of the plasma separatrix from the different codes is shown at a few shot times in Figure \ref{fig:lcfs_conv}. An excellent agreement can be observed between the codes.
% LCFS comparision
Following \cite{pentland2025validation}, absolute differences between the data from EFIT++ reconstruction and the three \gls{gs} solvers are reported in this paper. First, the poloidal flux on the axis (top panel) and the corresponding absolute differences (bottom panel) are shown in Figure \ref{fig:psia_conv}. The three codes are in a good agreement with an absolute difference below $10^{-2}$ Wb for the majority of shot duration.
% \psi_a comparision
Similarly, the poloidal flux at the plasma boundary is shown in Figure \ref{fig:psib_conv}. Here, the absolute difference between TokaMaker and EFIT++ is higher at about $10^{-2}$ Wb while it is about $10^{-3}$ Wb between EFIT++ and  Fiesta/FreeGSNKE.
% \psi_b comparision

\begin{figure}[hbt!]
 \centering
        \includegraphics[width=1.0\textwidth]{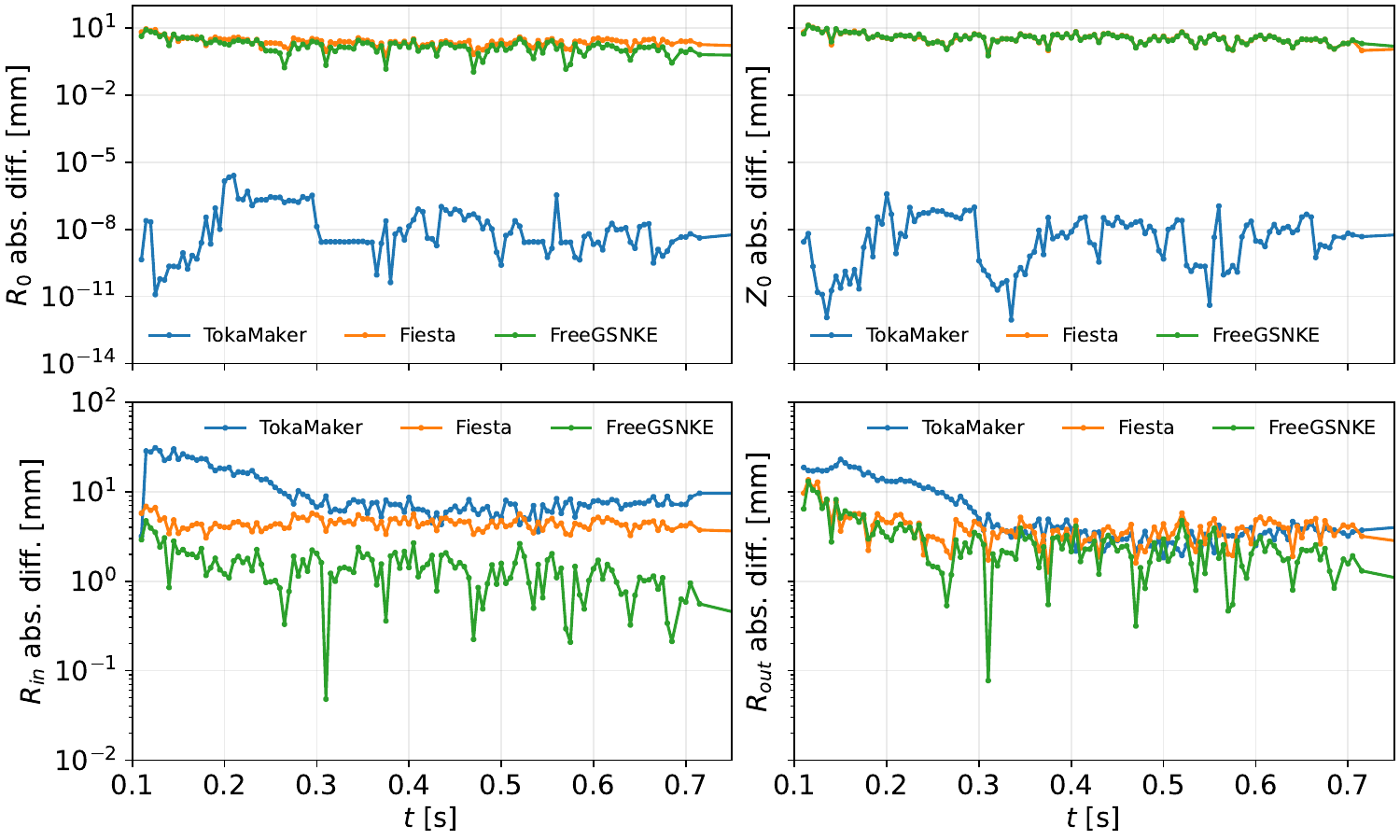}
 \caption{The absolute difference between EFIT++ and TokaMaker, Fiesta and FreeGSNKE for the magnetic axis components $R_0$ and $Z_0$ (top panel) and midplane inner $R_{in}$ and outer $R_{out}$ radii (bottom panel) in the conventional divertor configuration.}
\label{fig:r0_rin_conv}
\end{figure}

The absolute differences in shape targets such as the magnetic axis and the midplane extents of the sepratrix between EFIT++ and TokaMaker, Fiesta and FreeGSNKE are shown in Figure \ref{fig:r0_rin_conv}. The differences in the magnetic axis coordinates $R_0$ and $Z_0$ between EFIT++ and TokaMaker is close to the machine precision. The magnetic axis coordinates along with the plasma current $I_p$ are passed as hard constriants in TokaMaker which are satisfied exactly. The differences in the midplane inner radius $R_{in}$ and midplane outer radius $R_{out}$ are of the order of millimeters. The TokaMaker result of $R_{in}$ is slightly higher ($\sim 50$ mm) during the startup but steadily reduces to about 10 mm where it remains for the rest of the shot. The FreeGSNKE solution appears to be in a good agreement with the difference hovering around 1 mm for the majority of shot. A similar trend is observed for $R_{out}$, albeit with a lower abosulte difference. The TokaMaker result reaches a maximum value of about $\sim 25$ mm during the startup and gradually reaches a steady value of $\sim 5$ mm for the rest of the shot. The results from Fiesta and FreeGSNKE on the other hand have a lower value during the startup.

% R0_Rin comparision

\subsection{Super-X Divertor}
In the Super-X divertor configuration, the plasma at flat-top has a double-null configuration with a plasma current of about 750 kA. In this shot, no NBI heating was used and the plasma is entirely driven by Ohmic heating.

\begin{figure}[hbt!]
 \centering
        \includegraphics[width=1.0\textwidth]{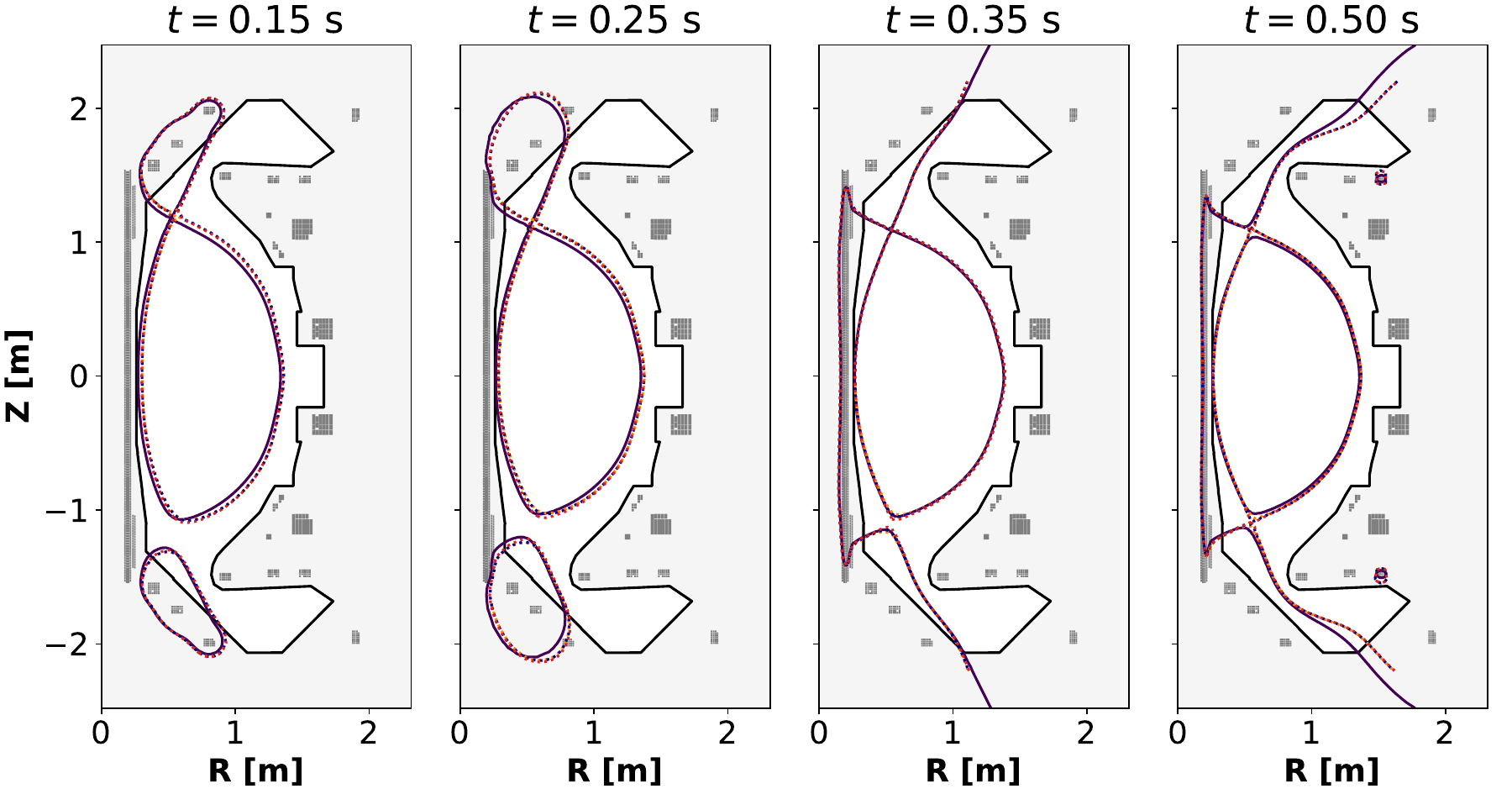}
 \caption{The evolution of TokaMaker (solid black), Fiesta (dotted orange), FreeGSNKE (dotted blue) and EFIT++ (doited red) separatrices at different times in the Super-X divertor configuration.}
\label{fig:lcfs_supx}
\end{figure}

\begin{figure}[hbt!]
 \centering
        \includegraphics[width=0.8\textwidth]{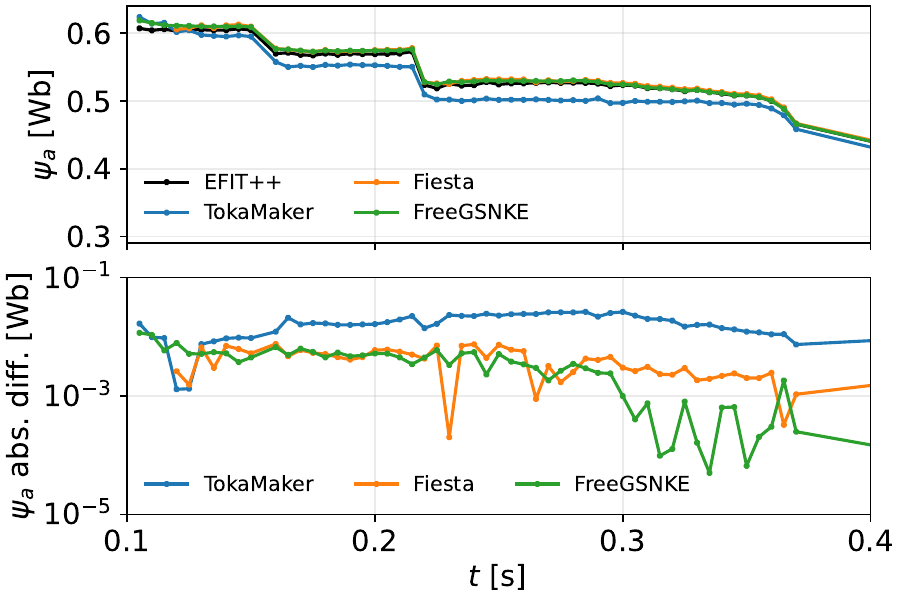}
 \caption{The evolution of poloidal flux at the magnetic axis of the plasma (top panel) and the absolute difference between EFIT++ and TokaMaker, Fiesta and FreeGSNKE (bottom panel) in the Super-X divertor configuration.}
\label{fig:psia_supx}
\end{figure}

\begin{figure}[hbt!]
 \centering
        \includegraphics[width=0.8\textwidth]{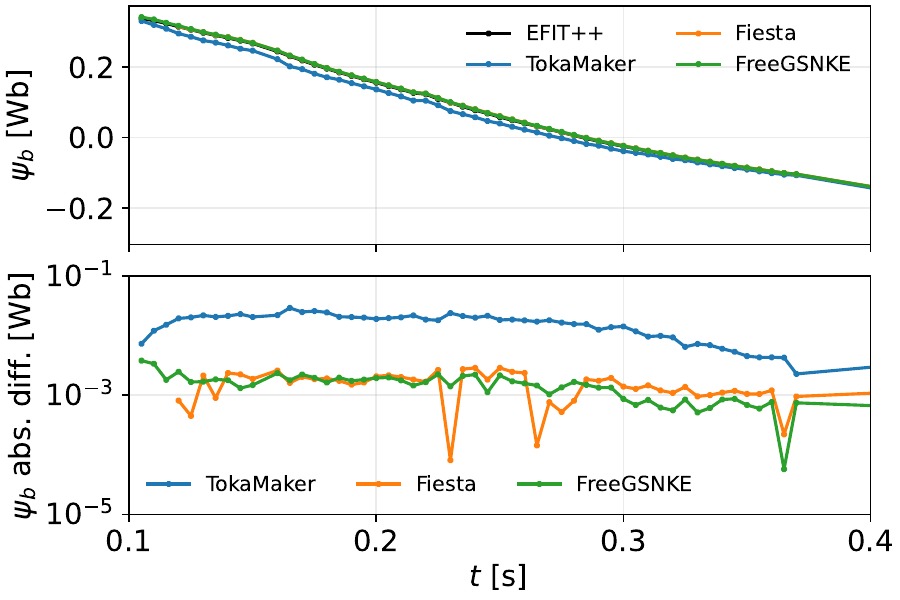}
 \caption{The evolution of poloidal flux at the plasma boundary (top panel) and the absolute difference between EFIT++ and TokaMaker, Fiesta and FreeGSNKE (bottom panel) in the Super-X divertor configuration.}
\label{fig:psib_supx}
\end{figure}

Just as was done for the conventional divertor configuration, the temporal evolution of the plasma separatrices from the different codes is shown in Figure \ref{fig:lcfs_supx}. Compared to the other codes, TokaMaker solution is slightly shifted at the early times but is in a good agreement at later times.
% At $t=0.5$ s, the solution from TokaMaker is a limited plasma while in the case of the other codes the plasma remains diverted.
% LCFS comparision
The poloidal flux on the axis (top panel) and the corresponding absolute differences (bottom panel) are shown in Figure \ref{fig:psia_supx}. The difference in the TokaMaker solution is higher than the other two codes but less than $10^{-1}$ Wb just as it was in the case of the conventional divertor configuration. Results from Fiesta and FreeGSNKE are below $10^{-2}$ Wb and in a better agreement with each other. A similar trend can be observed for $\psi_b$ as well in Figure \ref{fig:psib_supx} but with slightly lower absolute differences.

% \psi_a comparision

% \psi_b comparision

\begin{figure}[hbt!]
 \centering
        \includegraphics[width=1.0\textwidth]{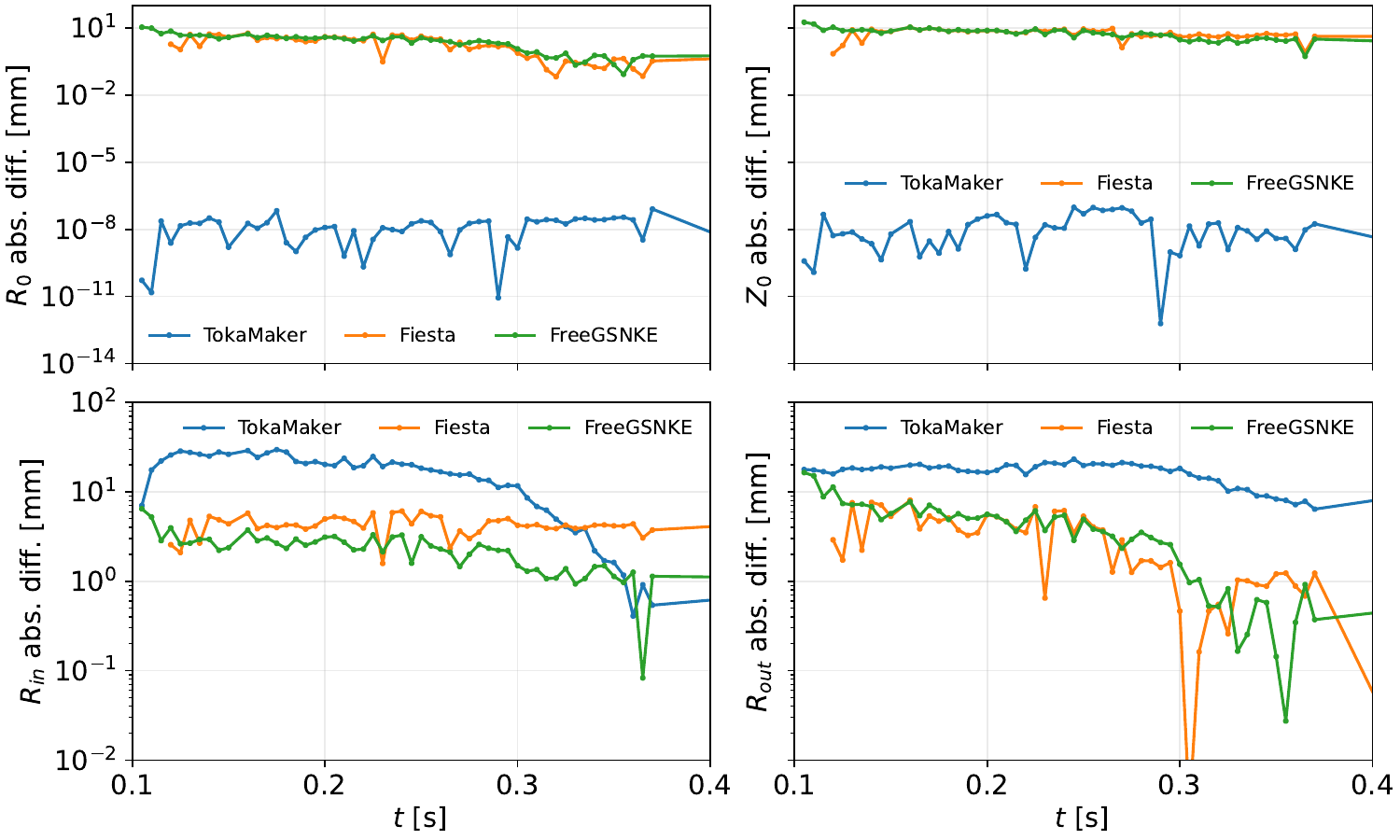}
 \caption{The absolute difference between EFIT++ and TokaMaker, Fiesta and FreeGSNKE for the magnetic axis components $R_0$ and $Z_0$ (top panel) and midplane inner $R_{in}$ and outer $R_{out}$ radii (bottom panel) in the Super-X divertor configuration.}
\label{fig:r0_rin_supx}
\end{figure}

Coming to the shape targets, a trend similar to the one observed in the case of the conventional divertor is observed. The temporal evolution of these parameters is shown in Figure \ref{fig:r0_rin_supx}. The magnetic axis is solved exactly in the case of TokaMaker while it differs by a 10 mm in the case of Fiesta and FreeGSNKE. The difference in $R_{in}$ and $R_{out}$ is greater than 10 mm in the case of TokaMaker while it is less than 5 mm in the case of Fiesta and FreeGSNKE.

% R0_Rin comparision

Overall, TokaMaker results are in a very good agreement with the EFIT++ reconstruction as well as Fiesta and FreeGSNKE solvers. The differences between TokaMaker results and results from the other \gls{gs} solvers can be attributed to the differences in the numerical methods, mesh sizes and the choice of \gls{vsc}.

% Neutronics
\section{Neutronics}\label{sec:neutronics}

Tokamaks for energy production are designed primarily to use \gls{dt} fusion~\cite{Ruhlig1938}. However, the handling of tritium (\ce{^{3}_{1}T}) is a complex process, not only because of its rarity, but also because of its radioactivity. Hence, most tokamaks begin their operations with \gls{dd} fuel, before undertaking an upgrade and moving onto using tritium as fuel. Consequently, a detailed neutronics study is crucial since both reaction pathways produce high energy neutrons.
The \gls{dt} reaction is an exothermic process that produces a helium nucleus, a neutron and \qty{17.6}{\MeV} energy, about 80\% of which manifest as the kinetic energy of the neutron; it is given by
\begin{equation*}
    \ce{^{2}_{1}D + ^{3}_{1}T -> ^{4}_{2}He + ^{1}_{0}n}.
\end{equation*}
The \gls{dd} fusion reaction has two pathways, once produces a tritium and a hydrogen nuclei, while the other gives rise to \qty{2.45}{\MeV} neutrons along with a helium-3 nucleus;
\begin{align*}
    \ce{^{2}_{1}D + ^{2}_{1}D &-> ^{3}_{1}T + ^{1}_{1}H},\\
    \ce{^{2}_{1}D + ^{2}_{1}D &-> ^{3}_{2}He + ^{1}_{0}n}.
\end{align*}
Neutronics study in a tokamak has three main modes of analysis. The focus areas in each mode are as follows:
\begin{enumerate}
    \item Mode 0: Magnet lifetime under neutron fluence, heat deposition in the first wall and blanket in addition to cooling systems. Maximize \gls{tbr} for long term self-sustained operations.
    \item Mode 1: \gls{sdr} for remote handling and maintenance.
    \item Mode 2: Decommissioning including categorization and storage of radioactive waste.
\end{enumerate}
For Mode 0, especially when designing a fusion relevant tokamak, some of the key takeaways from the study are the blanket size, radiation shield thickness, heat deposition and the corresponding required cooling systems.  

Once the operational parameters are defined (plasma parameters, temperatures, fuel) and the reactor is estimated to produce high energy neutrons, it is essential to estimate the neutron flux on the \gls{pfc}, \gls{fw}, blanket and radiation shield. Computing the neutron flux, involves solving a neutral particle transport equation, approximated by a neutron diffusion equation:
\begin{equation}
\begin{split}
    \left(\frac{1}{v(E)}\frac{\partial}{\partial t} + \hat{\Omega}\cdot\nabla + \Sigma_t(r, E, t)\right) &= \frac{\chi(r, E)}{4\pi}\int_0^\infty \text{d}E' \nu_p(r, E')\Sigma_f(r, E', t)\phi(r, E', t)\\
    &+ \int_{4\pi} \text{d}\Omega' \int_0^\infty \text{d}E' \Sigma_s(r, E' \rightarrow E, \hat{\Omega}' \rightarrow \hat{\Omega}, t)\psi(r, E', \hat{\Omega}', t)\\
    &+ s(r, E, \Omega, t).
\end{split}
\end{equation}
The independent variables are the position $r$, energy $E$, time $t$ and direction vector $\hat{\Omega} = \frac{\text{v}(E)}{|\text{v}(E)|}$ where $\text{v}$ is the neutron velocity. Here $\Omega$ is the solid angle in the direction of the velocity $\text{v}$ and the primed variables denote the ones to be summed over in the integration. The derived physical quantities are the angular neutron flux density $\psi$, neutron flux density integrated over the solid angle $\phi$, local neutron emissivity at a given energy $\nu(E)$, probability of neutron production $\chi$, total macroscopic cross section $\Sigma_t$, macroscopic cross section of fusion reaction $\Sigma_f$, differential scattering cross section $\Sigma_s$ and the source term $s$.

\begin{figure}[hbt!]
    \centering
    \includegraphics[width=\linewidth]{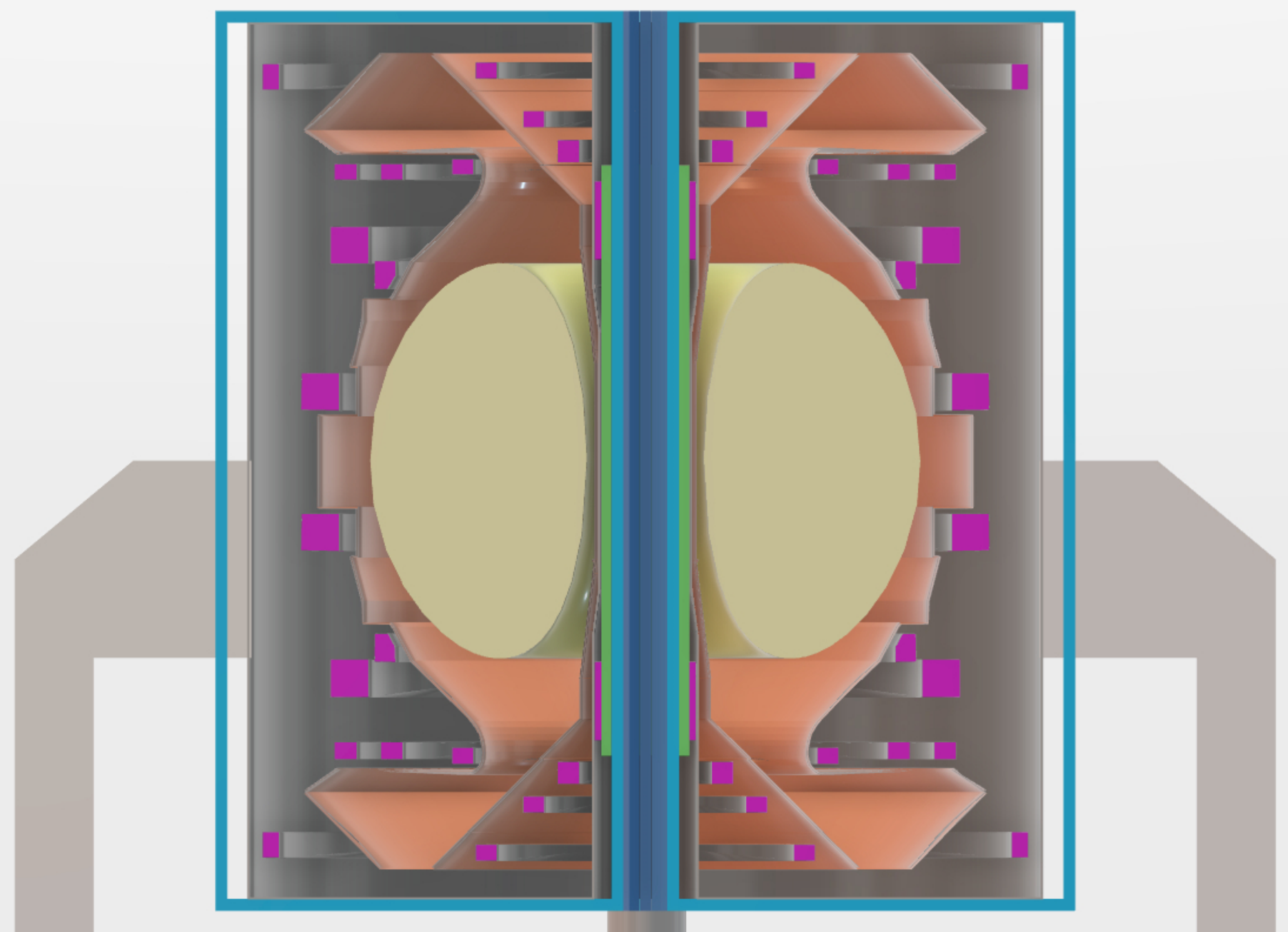}
    \caption{Cross-section of the \gls{cad} generated with Cadquery. The legs, vacuum vessel and \gls{pf} coil cases are stainless steel 304 shown in grey. The plasma, depicted in yellow, is a 50-50 mixture of deuterium and tritium. The conductors of the \gls{pf} coils are presented in pink, while those of the \gls{cs} are coloured green. The \gls{tf} coils are marked blue and the first wall is assigned orange. Any layer (mesh) which is not coloured is considered to be vacuum.}
    \label{fig:cad_model}
\end{figure}

Solving the above equation is challenging, since the solution is not only three dimensional with time dependence, but also spans a wide range of energies, \si{MeV} to sub-\si{eV}. Numerically solving this multi-dimensional equation with a high degree of fidelity requires enormous computational allocation, which is why stochastic methods are preferred. A Monte Carlo simulation of the above problem, while still challenging, can be solved without super-computing resources.
Numerical solutions are computed with stochastic codes like OpenMC~\cite{romano2015}, MCNP~\cite{kulesza2024}, TRIPOL-4~\cite{brun2014} and Serpent~\cite{leppanen2015} to name a few. In \jg, the OpenMC library/package is used for neutronics analysis.

\subsection{Workflow}

Setting up a neutronics study entails defining a geometry, specifying the materials, and adding the scattering cross sections for each pathway. In our framework, the logical steps are as follows:
\begin{enumerate}[label=(\roman*)]
    \item Create a reduced model \gls{cad} geometry with Cadquery~\cite{cadquery2024}, with assigned material tags for each solid geometry.
    \item The \gls{cad} model is imported using the \gls{dagmc}~\cite{wilson2010} framework. The \gls{dagmc} model is specified as the computational domain with the boundary conditions set to vacuum.
    \item Define material composition for each tag, including density and temperature.
    \item Specify the nuclear data library to use with the nulides specified in the materials list. \gls{fendl}~\cite{fendl2024}, \gls{endfb}~\cite{chadwick2011} or the \gls{jeff}~\cite{plompen2020} library is used.
    \item Define the source of the neutrons, which can be a point source, or a more realistic distributed source over the plasma cross section.
    \item Specify the tallies according to the requirements of the study. Here we focus on the flux and energy tally. We can also specify a particular reaction pathway (inelastic scattering $(n, n')$, neutron multiplication $(n, xn)$, tritium breeding $(n, nt)$, material activation $(n, \gamma)$) to use for the tally.
    \item Lastly, specify the settings for the neutron transport simulation, where at least the number of neutrons per batch and the number of batches must be specified.
\end{enumerate}

\subsection{Model}
A \gls{cad} geometry with reduced complexity is created using Cadquery from the
2D geometry available in the shot data. Here, all the physical build
specifications are parameterized. The 3D cross section of the \gls{cad} model is
shown in Figure \ref{fig:cad_model}. Every thickness and gap is parameterized
to accommodate for a greater degree of freedom for scoping alternative models.
For complex geometries, there is provision to use the outline in the poloidal
plane as input and create the solid with a specified thickness.

In this analysis, the geometry consists of three materials and the fuel of the
plasma. The vacuum vessel material is 304 Stainless Steel, the first wall material is graphite,
and the \gls{tf}, \gls{pf}, and \gls{cs} coils are pure copper. Although
\gls{mastu} is designed for \gls{dd} operations, in our study we study 
the neutron flux for an equivalent \gls{dt} operation. Hence, the plasma
material is defined to be a 50-50 mixture of \gls{dt}. This choice of materials
is motivated by \cite{reali2024}. The \gls{cad} geometry is imported with the
\gls{dagmc} framework into OpenMC as a reusable structure known as an
"universe". Here, we add a buffer space of \qty{0.02}{m} on all sides and define
the boundary as vacuum. Any gaps between two solid layers (for example the
plasma \gls{lcfs} and the first wall) are also filled with vacuum. Our 3D model
is a reduced version without many of the intricacies of the actual tokamak.

\subsection{Source}
The strength of the neutron source depends on the temperature and density of the ion. To this end, we impose an ion temperature profile and an ion density profile corresponding to an L-mode plasma and an assumed a neutron source density as outlined in \cite{bresard1991,fausser2012}: 
\begin{equation}
    S = S_0 \left( 1 - r^2/a^2 \right)^\lambda,
\end{equation}
where, $S_0$ is the central source density, $\lambda = 2\nu_n + \gamma\nu_T$ with $\nu_n$ and $\nu_T$ being the density and temperature profile peaking parameters respectively, $r$ is the radial location from the center of the plasma and $a$ is the minor radius of the plasma. The parameter $\gamma$ is taken from \cite{jarvis1994} which is taken to be $\gamma=3.5$ for $T_i= $~\qty{5}{keV} and $\gamma=2$ for $T_i=$~\qty{15}{keV}.
The central source density takes the form as given in \cite{wu2003}:
\begin{equation}
    S_0 = n_{i0}^2\langle\sigma v (T_{i0})\rangle,
\end{equation}
where $n_{i0}$ and $T_{i0}$ are the density and temperature of the central ion respectively.

\begin{figure}[hbt!]
    \centering
    \includegraphics[width=\linewidth]{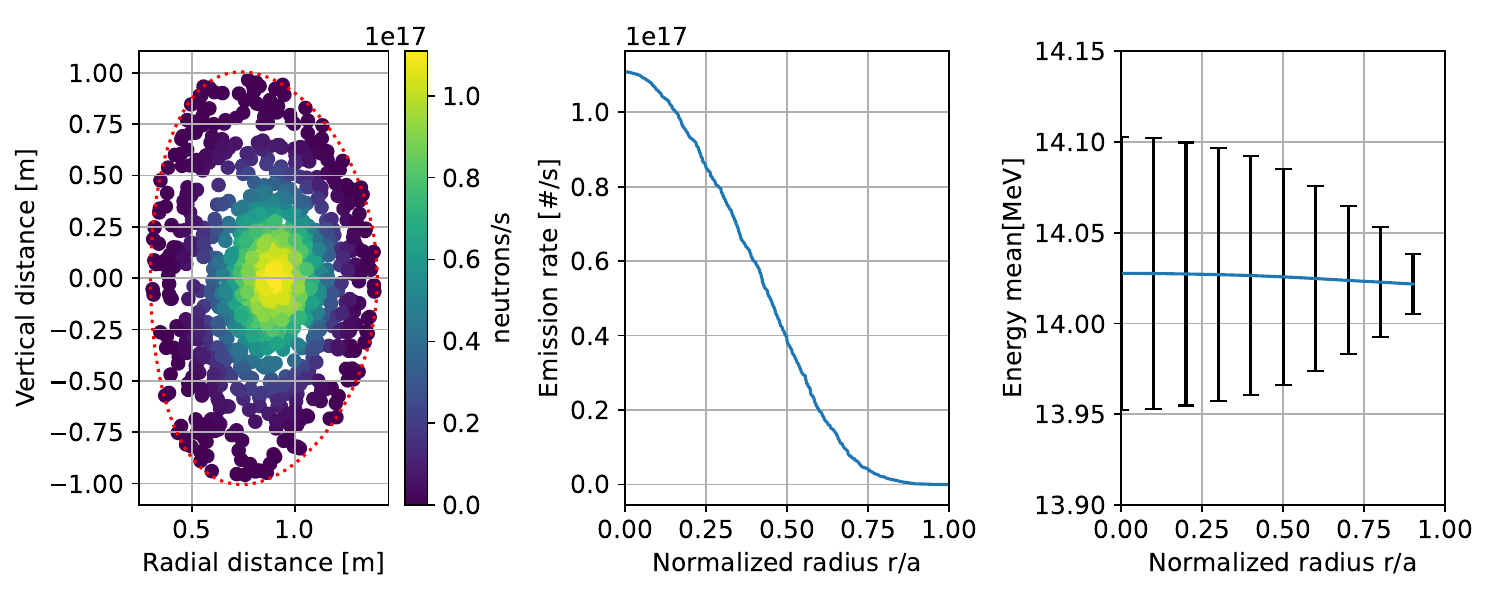}
    \caption{The (left) panel shows the distribution of the neutron source over the region bounded by the \gls{lcfs} (dotted red). The plot is in the poloidal (R, Z) plane and the colorbar represents the neutron source density. The (center) panel shows the radial neutron source distribution, normalized to the minor radius $a$. The (right) panel shows the average energy of the neutrons over the normalized radius, while the vertical errorbars represent one standard deviation of the neutron energy distribution along each ring source located at that radius.}
    \label{fig:source_dist}
\end{figure}

The energy of the neutrons also depends on the temperature of the ions. We define the mean and standard deviation of the neutron energy distribution using the form given in \cite{ballabio1998}. This explicitly breaks the radial symmetry of neutron energy via its dependence on the ion temperature, which varies over the radius. The first and second moments are represented in terms of the noralized radius $\rho$, as in~\cite{ballabio1998}:
\begin{equation}
    \begin{split}
        \langle E\rangle &= E_0 + \Delta E(\rho),\\
        \sigma &= \frac{\omega_0 (1 + \delta)\sqrt{T_i(\rho)}}{2\sqrt{2 \ln{2}}},
    \end{split}
\end{equation}
where $E_0$ is the mean energy, $\Delta E$ is the correction to the first moment, $T_i$ is the ion temperature, $\delta$ is the correction term, $\omega_0$ is the uncorrected standard deviation.
The correction term for the first moment is defined as:
\begin{equation}
     \Delta E(\rho) = \frac{\alpha_1}{1 + \alpha_2 T_i(\rho)^{\alpha_3}} T_i(\rho)^{2/3} + \alpha_4 T_i(\rho).
\end{equation}

 The~\gls{mhd} equilibrium lines lie within the isobaric surfaces, which are also isothermal, implying iso-source density. Hence, provided that the poloidal diffusion and convection do not change the iso-flux magnetic surfaces significantly, we can assume that the neutron source density and energy distributions are fixed and bound by the \gls{lcfs}. Following \cite{chen2003}, the \gls{lcfs} is given by
\begin{equation}
    \begin{split}
    R &= R_0 + a\cos(\theta + \delta\sin(\theta)) + \xi\left(1 - r^2/a^2 \right) \\
    Z &= \kappa a \sin(\theta),
\end{split}
\end{equation}
where, $R_0$ and $a$ are the major and minor radius of the plasma respectively, $r$ is the radial location from the centroid of the plasma, $\theta$ is the poloidal angle, $\delta$ is the triangularity, $\kappa$ is the elongation and $\xi$ represents the Shafranov shift. This is simply a co-ordinate transformation from $(r, \theta) \rightarrow (R, Z)$.

\begin{figure}[hbt!]
    \centering
    \includegraphics[width=0.8\textwidth]{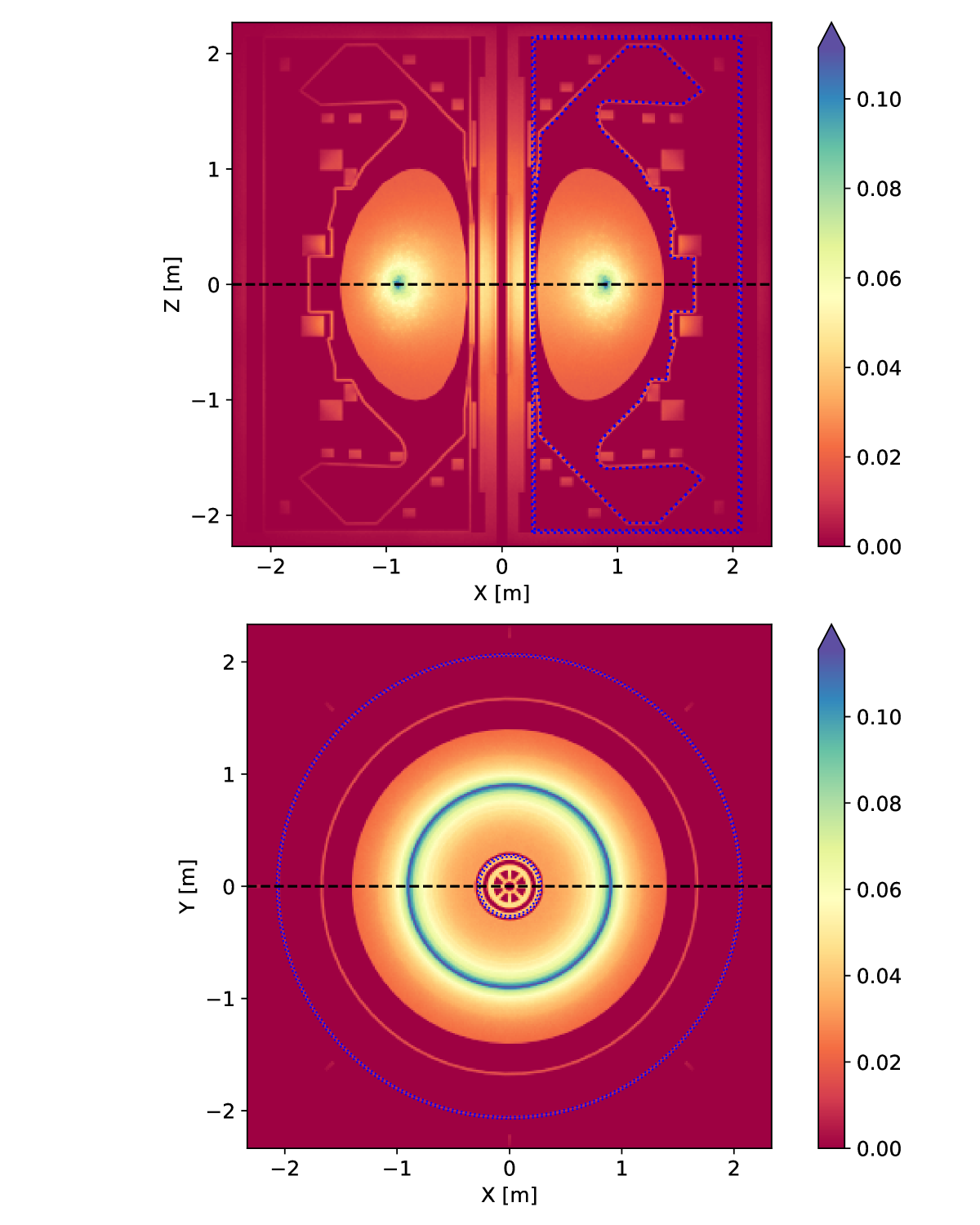}
    \caption{Cross section of the tokamak along the XZ plane (top) and the XY (bottom) plane, with the colorbar representing the neutron flux in units of neutrons cm per source. The neutrons are tallied only at the solid geometries, which are visible as the lighter regions in the \gls{cad}. The elements are added via mesh filters for the flux tally. The plots outline the shape of the first wall and the vacuum vessel with a blue outline.}
    \label{fig:flux_cross_section}
\end{figure}

% \subsection{Settings}
\subsection{Results}
In the following section, we summarize the results of our OpenMC study of a conceptual \gls{dt} operation scenario in the MAST-U tokamak. In this study, a total of $10^7$ neutrons per batch was used, averaged over $20$ batches. The computational domain was discretized into a uniform mesh of size $200^3$ with vacuum boundary conditions. All simulations were carried out on a workstation running on a Ryzen 9900X processor with $24$ threads and $64$~GB of DDR5 SDRAM. Memory proved to be a bottleneck, as the volume mesh generated by OpenMC scales as $O(N^3)$, with added space required for I/O (writing the state file and tallies to disc). The number of filters for specific tallying is a post-processing problem limited by the I/O capabilities. Neutron trajectories are computed in parallel using \gls{mpi}. The study is motivated by a similar study done with the OpenMC library in \cite{reali2024}, wherein a pre-made full \gls{cad} was used for the analysis.

\begin{figure}[hbt!]
    \centering
    \includegraphics[width=0.8\linewidth]{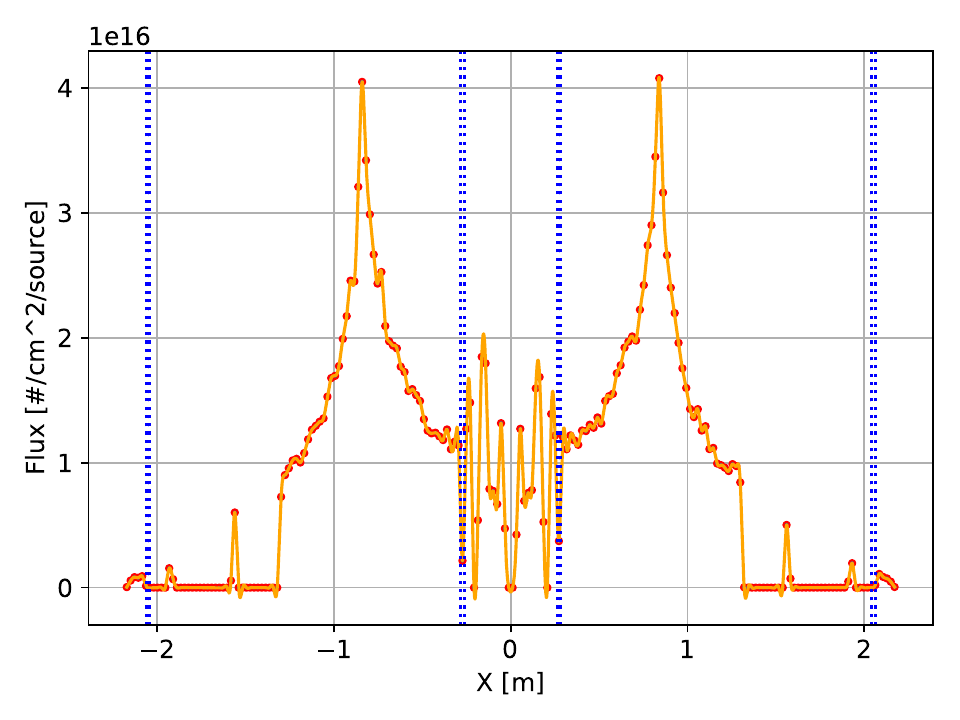}
    \caption{Representation of the 1D slice of the spatial flux distribution, along the central line denoted in Figure \ref{fig:flux_cross_section} with a black dashed line. The blue dashed lines correspond to the same in Figure \ref{fig:flux_cross_section}. The 1D plot gives a qualitative idea of how the neutron flux varies radially in a poloidal plane.}
    \label{fig:flux_linear_section}
\end{figure}

% \subsection{Results}
Following \cite{fausser2012}, the neutron source is initialized as a collection of isotropic ring sources with azimuthal symmetry. The distribution of the neutron source density over the poloidal plane of the plasma is shown in Figure \ref{fig:source_dist}. They are bounded by the plasma \gls{lcfs}, marked by the red dotted line on the left subplot. Within the \gls{lcfs}, the temperature of the ions is above the threshold for the \gls{dt} reaction. The colorbar of each ring source point denotes its source density. The radial source distribution in the center plot depends on the peaking parameters of the density and temperature profiles. The energy of the neutrons emitted from a single ring source follows a normal distribution, whose mean and standard deviation vary radially. The subplot on the right shows the mean as the blue curve and the one standard deviation as the error bars. 

Figure \ref{fig:flux_cross_section} shows the cross sections of the full 3D geometry along the ZX (top) and XY (bottom) planes, respectively, where the colorbars represent the average neutron flux in units of [neutrons~\si{cm}/source]. In OpenMC the flux is computed per voxel, after averaging over the number of source particles and batches. We have also outlined the inner boundary of the first wall and the section that defines the vacuum vessel. As expected, the neutron flux is highest at the center of the plasma, where the source density is highest and dampens radially outward.

\begin{figure}[hbt!]
    \centering
    \includegraphics[width=\linewidth]{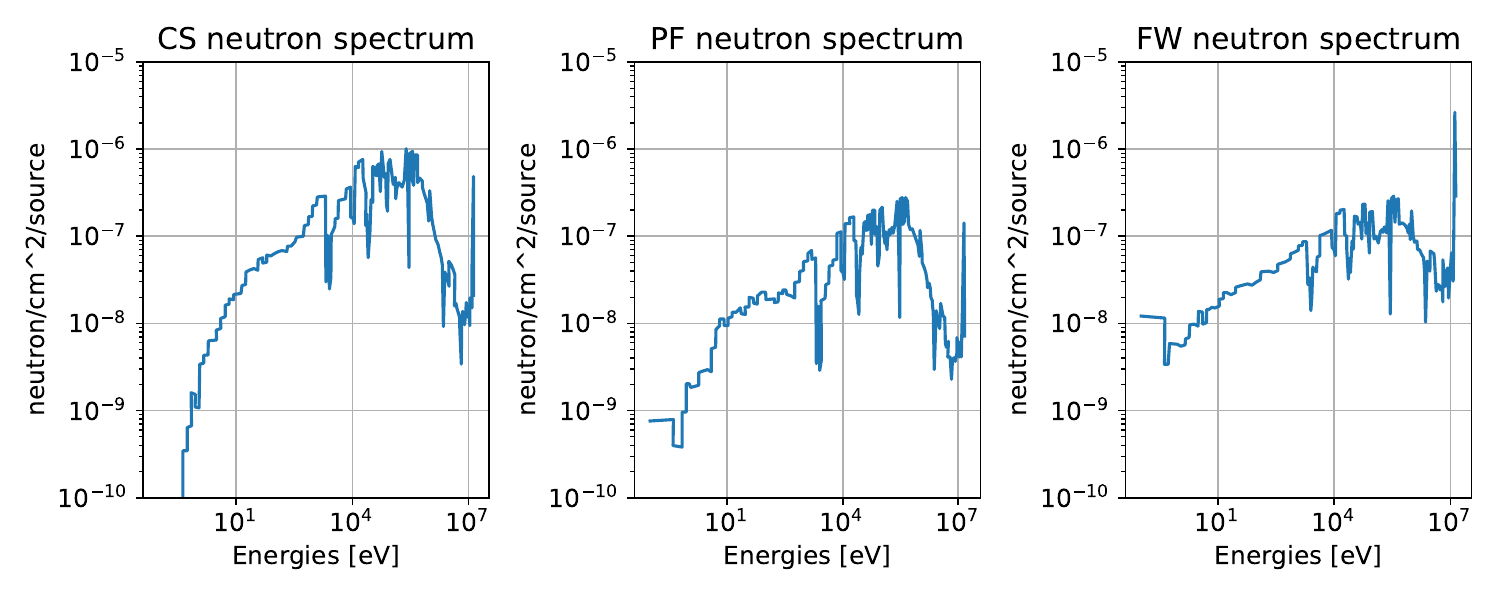}
    \caption{The results from \cite{reali2024}, which uses a detailed MAST-U \gls{cad}, are reproduced. The x-axis and y-axis represent the energy in \si{eV} and the flux in units of neutrons/\si{cm^2}/source respectively.}
    \label{fig:comparison_data}
\end{figure}

In Figure \ref{fig:flux_linear_section} the plot represents the 1D distribution of the flux along the horizontal dashed line drawn in Figure \ref{fig:flux_cross_section}. The vacuum vessel boundary is marked with vertical blue dashed lines. We get a qualitative idea of the neutron flux distribution as we move radially out from the axis of the tokamak, with the perspective of the thicknesses of each layers. Both the cross sections and the linear distribution are plotted over the diameter and height of the tokamak, with some added padding on all sides.

In order to visualize the distribution of neutron flux over the energy spectrum in each material, we have presented the material specific flux spectra in our analysis plots. For each region that needs to be tallied, we add a mesh filter with the given material tag. Since we are summing over volume and only segregating on the basis of energy of the neutrons, the mean flux needs to be normalized by the volume of the respective mesh. To this end, we computed the volume of the tallied meshes, using a stochastic method with the OpenMC package, for normalisation. When reporting flux tallies, OpenMC uses the units neutron-cm/source, which we divide by the mesh volume and the source amplitude to get neutrons/\si{cm^2}/source. For the neutron energy spectra, we use a bin size of $200$. 

\begin{figure}[hbt!]
    \centering
    \includegraphics[width=\textwidth]{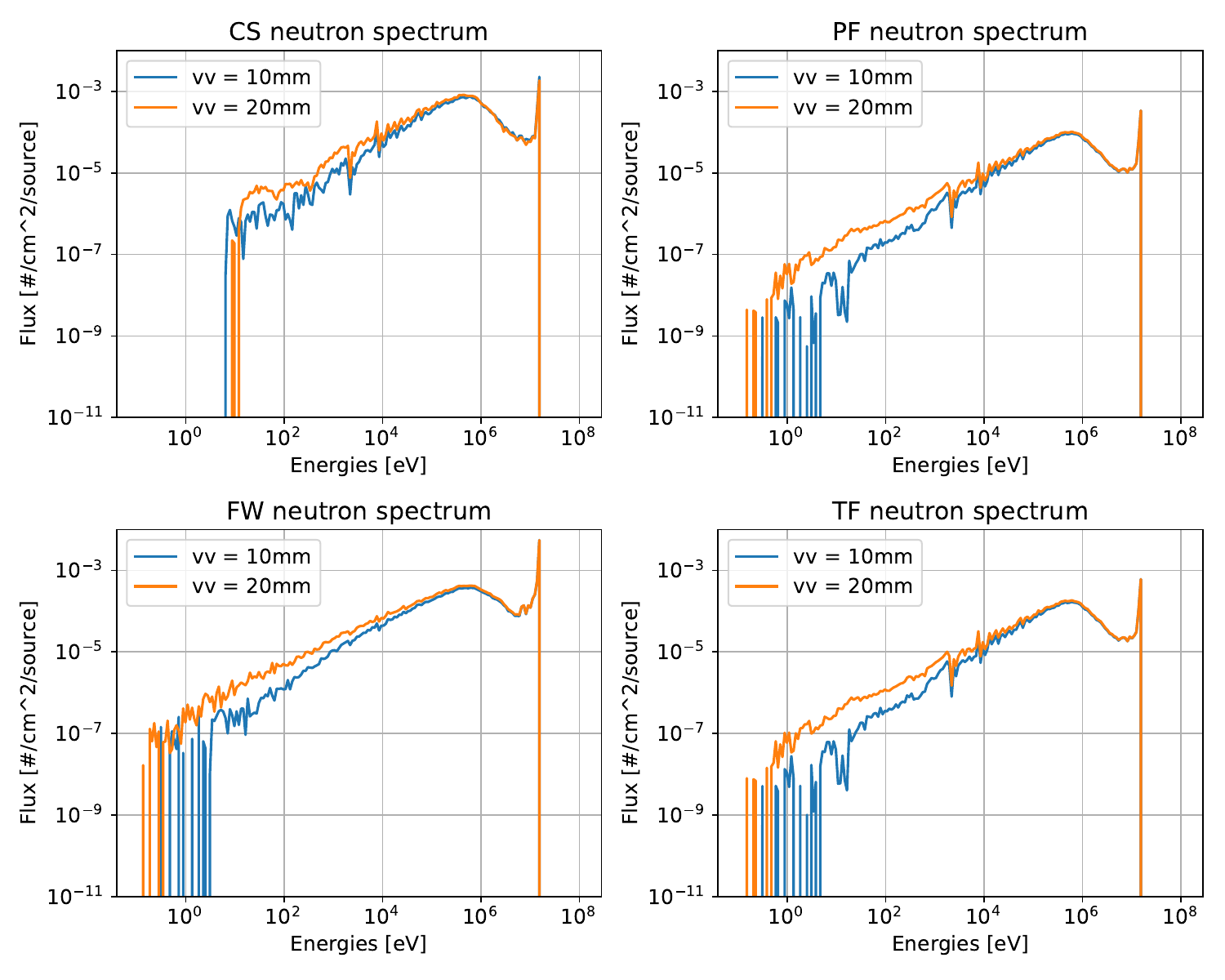}
    \caption{Neutron flux spectra in each individual layer, tallied over $200$ bins, spanning over the entire range of energies. The x-axis represents energy in \si{eV} and the y-axis shows the flux in units of neutrons/\si{cm^2}/source. The \gls{cs}, \gls{pf}, \gls{fw} and \gls{tf} material filters are used for the respective plots. There was no corresponding data for \gls{tf} in~\cite{reali2024}.}
    \label{fig:flux_spectra_fw10}
\end{figure}

\begin{figure}[hbt!]
    \centering
    \includegraphics[width=\textwidth]{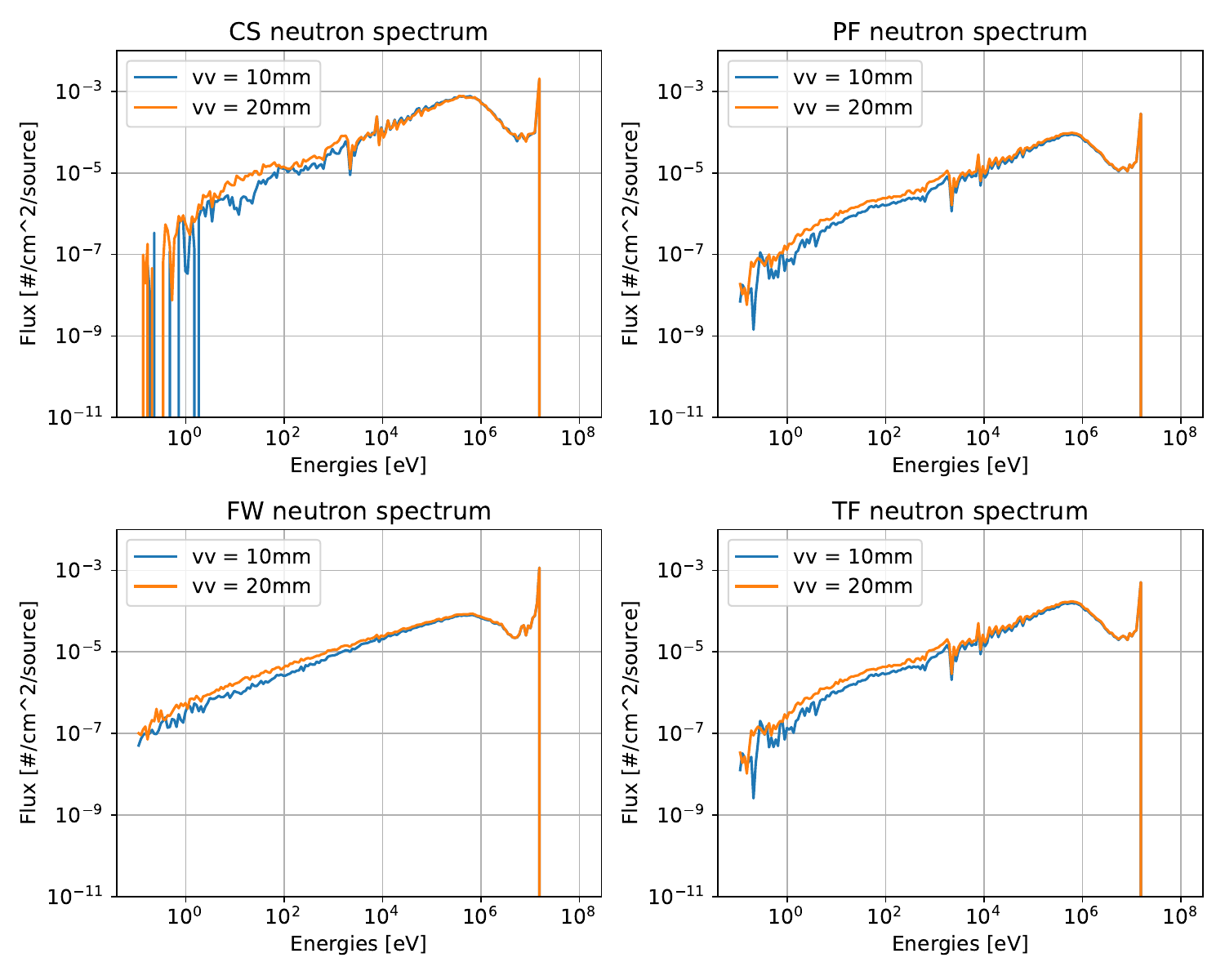}
    \caption{Neutron flux spectra in each individual layer, tallied over $200$ bins, spanning over the entire range of energies. The x-axis represents energy in \si{eV} and the y-axis shows the flux in units of neutrons/\si{cm^2}/source. The \gls{cs}, \gls{pf}, \gls{fw} and \gls{tf} material filters are used for the respective plots. There was no corresponding data for \gls{tf} in~\cite{reali2024}.}
    \label{fig:flux_spectra_fw20}
\end{figure}

The design of the first wall and the vacuum vessel in \gls{mastu} are modular, composed of many cylindrical tiles with varying thickness in the radial (and vertical) dimensions. In our study, we worked with a reduced order \gls{cad} model, where we maintained a parameterized fixed width for each of these layers. In order to cover the limiting cases, we performed our analysis for two first wall thicknesses and two vessel thicknesses (\qty{10}{mm} and \qty{20}{mm}). We compare the fluxes across the \gls{cs}, \gls{pf}, \gls{fw} and \gls{tf} for first wall thicknesses of \qty{10}{mm} in Figure \ref{fig:flux_spectra_fw10} and \qty{20}{mm} in Figure \ref{fig:flux_spectra_fw20}. In each of these figures, we have compared two models representing the aforementioned vacuum vessel thicknesses.

The sharp peak near \qty{14}{MeV} marks the energy of the source neutrons in the material. This peak is highest for the \gls{fw}, since it is the first layer encountered by the source neutrons beyond the \gls{lcfs}. As we move radially outwards, this central peak dampens, going from the \gls{fw} to \gls{cs} to \gls{tf} to \gls{pf}. Although the \gls{fw} is closest to the neutron source, the extensions for the divertor regions reduce the average flux per surface area of low and intermediate energy neutrons, due to an increase in its total volume. The central solenoid is subject to the highest energy flux, making it more critical in the design of a compact \gls{st}.

Increasing the thickness of either of the two layers has two pronounced effects: a decrease in the source neutron flux (sharp peak around \qty{14}{MeV}) and an increase in the flux of low and intermediate energy neutrons (more prominent in low energies). The graphite moderating the neutrons via inelastic scattering pathways is responsible for this behaviour. Increasing the thickness of the vacuum vessel shows a similar trend with the low energy neutrons showing a higher flux, indicating inelastic scattering.
Compared to the results from \cite{reali2024} in Figure \ref{fig:comparison_data}, a larger flux of high energy neutrons is observed across all the layers. However, the trend remains consistent across the two studies. This is attributed to the simplified \gls{cad} model used in our work which has 60 independent components compared to 3938 individual component in \cite{reali2024}.
% In our comparison with the work presented in \cite{reali2024}, it is clear that we have a larger flux of high energy neutrons across layers compared to Fig.~\ref{fig:comparison_data}, due to the reduced order model with fewer solid objects, which become prohibitive to generate parametrically. The general shape is commensurate with our results.

Designing a new shield must make such a study over a wide range of materials of varying thickness, which can be easily set up using \jg, owing to its parametric construction. We specify the materials for each layer with a separate dataclass, where the options are indexed by integers. It is easy to set up a batch simulation for a whole range of materials, including similar materials with slightly different fractions or isotope enrichments.
One point of note is that there are regions showing zero neutron flux (dark red) between the plasma~\gls{lcfs} and the first wall, which are not physical, but is an artifact of the set up. The \gls{cad} is built to have a solid that defines the plasma followed by another solid that defines the first wall, with nothing in between. We set up OpenMC such that any region that is not tagged with a material automatically gets assigned to be a ``vacuum''. This can be addressed by adding a neutral gas region with the appropriate density. The neutron flux and the energy flux spectra are both strongly dependent on the size of the voxels or inversely, the mesh size for the neutronics simuations. Having a large voxel (bin) leads to much higher fluxes than what would be detected by a liquid scintillator~\cite{morgan1975}. Usually the scintillators have a dimension of approximately \qty{5}{cm} in diameter and length, implying that an accurate simulation, that correlates with the neutron flux measurements from the detector, would require a mesh size such that each voxel is at most a cube of size~\qty{5}{cm}. Given the dimension of the \gls{mastu}, the voxel size is $2.34 \times 2.34 \times 2.27$~\si{cm^3} for the $200^3$ lattice in the simulation. The side of an approximate cubic voxel of the same volume is half of the characteristic size of the detector. Going to a finer mesh would give a lower and more ``accurate'' flux, but that is meaningless unless it can be measured by a detector in an experimental setup.

\section{Conclusions}
In this work, we introduce \jg\ which seamlessly integrates 0D scaling relations used in system studies to 2D/3D first-principle models in a unified and modular framework. The operation of \jg\ was demonstrated by modeling the \gls{mhd} equilibrium of \gls{mastu} discharges using a forward \gls{gs} solve and a neutronics study to estimate the neutron irradiation of various structures in MAST-U for a \gls{dt} operation.

The EFIT++ reconstruction data from \cite{pentland2025validation} was used to setup the \gls{mastu} geometry which is comprised of the locations and currents in the active coils and the passive conductors, plasma profiles ($p'$, $FF'$) and magnetic axis. This information along with numerical parameters specific to the \gls{gs} solver were used in the forward solve for the entire pulse duration. A fictitious \gls{pf} coil was introduced to stabilize the plasma against vertical instability and to match all the active coil currents to match that in the EFIT++ reconstruction. A good agreement between \jg\, EFIT++, Fiesta and FreeGSNKE was obtained for several quantities such as the shape targets (radii of the midplane), magnetic axis, separatrices, poloidal flux on the plasma axis and \gls{lcfs}. The differences in the shape targets parameters were observed to be within 10 cm and that for the flux quantities were well below 0.01 Wb for most of the time slices. The \gls{gs} solver, previously validated against analytic equilibria \cite{hansen2024109111}, has now been validated against experimental reconstruction and cross-verified with established \gls{gs} solvers Fiesta and FreeGSNKE.

We generated a reduced order \gls{cad} in real time using a parametric model, which was used as the solid geometry for neutron transport. It was composed of three primary materials and a distributed neutron source bounded by the \gls{lcfs}. The source density is modeled by a radial gaussian with the energy set by the ion density and temperature. We tallied both the total flux and the partial flux in components, namely the coils and the \gls{fw}. The total flux gave us a representation of the spatial flux distribution. The partial one, although volume averaged, provides information about the essential moderation processes and how the layer thickness changes neutron flux across energy spectrum. The parametric reduced-order geometry, although with orders or magnitude fewer individual components and surfaces, could qualitatively match the results of a similar work \cite{reali2024}. This entire workflow from generating a geometry to computing the partial flux spectra is done within the framework of \jg, while being able to take design decisions (like changing the thickness of the vacuum vessel) on the fly for quick prototyping.

The \jg\ framework is under active development to couple transport to provide a more detailed temperature and density profiles, a plasma control system for designing and prototyping actuator waveforms, uncertainty quantification, real-time plasma reconstruction and integrating the IMAS data structure.
% \clearpage

% \ack{Sample text inserted for demonstration.}

% \funding{Sample text inserted for demonstration.}
% % This section is a list of funder names and grant numbers

% \roles{Sample text inserted for demonstration.}
% % List author names and the contributions made to the article, using terms from the NISO Contributor Roles Taxonomy (CRediT) https://credit.niso.org

% \data{Sample text inserted for demonstration.}
% % For more information on IOP Publishing's research data policy see: https://publishingsupport.iopscience.iop.org/questions/research-data/

% \suppdata{Sample text inserted for demonstration.}

%--------------------------References----------------------------
% \section*{References}   %<--- Provided by biblatex
\printbibliography
% \bibliographystyle{apa}
% \bibliography{refs}

\end{document}